\documentclass[twocolumn]{aastex62}

\usepackage{apjfonts}
\usepackage{subfigure}
\usepackage{amsmath}
\usepackage{xspace}
\usepackage{hyperref}

\newcommand{\kr}{$\kappa_{\mathrm{R}}$\xspace}
\newcommand{\amax}{$a_{\mathrm{max}}$\xspace}
\newcommand{\tpop}{\texttt{twopoppy}\xspace}
\newcommand{\vfrag}{$v_{\mathrm{frag}}$\xspace}
\newcommand{\alphat}{$\alpha_{\rm t}$\xspace}

\shorttitle{Radial Gradients in Dust-to-Gas Ratio}
\shortauthors{Chachan, Lee, and Knutson}

\begin{document}

\title{\textbf{\large{Radial Gradients in Dust-to-Gas Ratio Lead to Preferred Region for Giant Planet Formation}}}
\correspondingauthor{Yayaati Chachan}
\email{ychachan@caltech.edu}

\author[0000-0003-1728-8269]{Yayaati Chachan}
\affil{Division of Geological and Planetary Sciences, California Institute of Technology, 1200 E California Blvd, Pasadena, CA, 91125, USA}

\author[0000-0002-1228-9820]{Eve J. Lee}
\affiliation{Department of Physics and McGill Space Institute, McGill University, 3550 rue University, Montr\'eal, QC, H3A 2T8, Canada}
\affiliation{Institute for Research on Exoplanets, Montr\'eal, QC, Canada}

\author[0000-0002-5375-4725]{Heather A. Knutson}
\affil{Division of Geological and Planetary Sciences, California Institute of Technology, 1200 E California Blvd, Pasadena, CA, 91125, USA}

\begin{abstract}

The Rosseland mean opacity of dust in protoplanetary disks is often calculated assuming the interstellar medium (ISM) size distribution and a constant dust-to-gas ratio. However, the dust size distribution and the dust-to-gas ratio in protoplanetary disks are distinct from those of the ISM. Here, we use simple dust evolution models that incorporate grain growth and transport to calculate the time evolution of mean opacity of dust grains as a function of distance from the star. Dust dynamics and size distribution are sensitive to the assumed value of the turbulence strength \alphat and the velocity at which grains fragment \vfrag. For moderate-to-low turbulence strengths of $\alpha_{\mathrm{t}} \lesssim 10^{-3}$ and substantial differences in \vfrag for icy and ice-free grains, we find a spatially non-uniform dust-to-gas ratio and grain size distribution that deviate significantly from the ISM values, in agreement with previous studies. The effect of a non-uniform dust-to-gas ratio on the Rosseland mean opacity dominates over that of the size distribution. This spatially varying---that is non-monotonic---dust-to-gas ratio creates a region in the protoplanetary disk that is optimal for producing hydrogen-rich planets, potentially explaining the apparent peak in gas giant planet occurrence rate at intermediate distances. The enhanced dust-to-gas ratio within the ice line also suppresses gas accretion rates onto sub-Neptune cores, thus stifling their tendency to undergo runaway gas accretion within disk lifetimes. Finally, our work corroborates the idea that low mass cores with large primordial gaseous envelopes (`super-puffs') originate beyond the ice line.

\end{abstract}

\section{Introduction}

Dust opacity plays an important role in setting the temperatures and vertical structures of protoplanetary disks \citep[e.g.,][]{Chiang1997, DAlessio1998} and determines how rapidly a planet accretes its gaseous envelope \citep[e.g.,][]{Stevenson1982a, Pollack1996, Ikoma2000}. The temperature structure of the disk determines where various molecules can condense, resulting in a spatially and temporally varying division of elements between solid and gas phases \citep[e.g.][]{Hayashi1981, Oberg2011}. In the core accretion framework, dust opacity regulates the cooling of the envelope accreted by a growing planet \citep[e.g.,][]{Piso2014, Lee2014, Piso2015}. Because the envelope accretion rate is cooling-limited during the hydrostatic phase of planetary growth, this dust opacity also has a strong influence on the final envelope mass.

In particular, cooling-limited accretion determines which planetary cores reach the threshold for runaway gas accretion within the gas disk lifetime and hence influences the giant planet occurrence rate. Radial velocity surveys indicate that giant planets inside 7 au only occur around 10\% of FGK stars and they predominantly orbit their host stars at intermediate distances ($3-5$ au); their occurrence rate declines at both smaller and larger orbital distances (\citealp{Cumming2008, Howard2012, Wittenmyer2016, Fernandes2019, Wittenmyer2020, Rosenthal2021, Fulton2021}; complemented by direct imaging surveys, e.g. \citealp{Bowler2018, Baron2019}). It is unclear why giant planets preferably occur at intermediate distances. The water ice line is typically assumed to play a role in making this region favorable for giant planet formation, primarily by facilitating the formation of massive cores \citep[e.g.][]{Morbidelli2015}. However, the role of gas accretion in shaping the occurrence rate of giant planets remains largely unexplored.

Sub-Neptunes dominate the observed population of exoplanets with orbital periods less than 300 days \citep[e.g.,][]{Batalha2013, Fressin2013, Morton2014, Dressing2015, Petigura2018}. The measured radii and masses of sub-Neptunes are consistent with hydrogen and helium  envelope mass fractions of a few percent \citep{Wolfgang2015, Ning2018}, despite the fact that some of these planets have cores massive enough ($\gtrsim$10$M_\oplus$) to reach the threshold for runaway gas accretion. What regulates the envelope mass fraction at a few percent? It has been suggested that the accretion of material with high dust opacity could prevent these planets from amassing significantly larger envelopes \citep[e.g.][]{Lee2014, Chen2020}. Here, we revisit this idea and explore why sub-Neptunes might be expected to form close-in whereas gas giants are more common at larger orbital separations.

Determining the dust opacity at a given location in the protoplanetary disk is a non-trivial task as it depends on the poorly known optical properties (composition and structure), size distribution, and dust-to-gas ratio, all three of which are intricately coupled to the protoplanetary disk's structure and evolution. Previous studies in both the protoplanetary disk and planet formation literature \citep[e.g.][]{Bell1994, Alexander1994} have generally elected to adopt a single global value for the dust-to-gas ratio and a power-law size distribution (both the power-law index and the bounding grain sizes) that is akin to that of dust in the interstellar medium (ISM). However, such a prescription is too simplistic to calculate the mean opacity due to dust grains in a protoplanetary disk. Dust grains in protoplanetary disks grow to sizes that are significantly larger \citep[mm-cm size, e.g.][]{Miyake1993, Testi2003, Draine2006, Andrews2015} than the largest sub-micron sized grains in the ISM \citep{Draine1984}. A larger maximum grain size redistributes dust mass from smaller grains to larger grains, which significantly alters the short-wavelength and mean opacities of protoplanetary disks \citep[e.g.][]{DAlessio2001}.

Fortunately, advances in our understanding of grain coagulation and the role of fragmentation and radial drift in limiting grain growth now make it possible to calculate the grain size distribution as a function of location in protoplanetary disks \citep{Brauer2008, Birnstiel2010, Birnstiel2011}. In a recent study, \citet{Savvidou2020} assessed the effect of varying grain size distribution from coagulation and fragmentation on the Rosseland mean opacity and the thermal structure of the disk, but without taking dust transport into account. Transport of dust due to radial drift, gas drag, and turbulent diffusion leads to a radially-varying dust-to-gas ratio, which may significantly alter dust opacity. This in turn leads to a location dependence of the gas accretion rates onto planetary cores. The gas accretion rate depends on the envelope's ability to cool at the innermost radiative-convective boundary (RCB). For envelopes in which dust opacity dominates, the sublimation of dust leads to the formation of an intermediate radiative zone and the innermost RCB is set by the H$_2$ dissociation front \citep{Lee2014}. Cooling at the innermost RCB is then controlled by H$^-$ opacity, which depends on the availability of free electrons provided by metals. These metals are mostly present in the dust initially and so the radially-varying dust-to-gas ratio of the accreted material affects the rate at which the envelope can cool.

In this work, we use a published dust evolution model to calculate the spatial and temporal evolution of the dust-to-gas ratio in a protoplanetary disk \citep{Birnstiel2012} in \S~\ref{sec:modelling}.  We then calculate the corresponding Rosseland mean opacity using an approximate size distribution scheme to determine the grain size distribution as a function of distance from the star \citep{Birnstiel2015}. In \S~\ref{sec:results}, we compute the disk opacity from dust evolution models as a function of radial distance, height from midplane, and time and show that our results differ starkly from the usual ISM opacity values. We then use our updated opacity values to calculate gas accretion rates onto planetary cores using the analytical scaling laws from \cite{Lee2015} 
and discuss the consequences of our work for the formation of gas giants, sub-Neptunes, and `super-puffs' (low mass planets with sizes beyond $\sim4 \,$R$_\oplus$) in \S~\ref{sec:planet_formation}. We summarize our results and suggest potential directions for future work in \S~\ref{sec:conclusion}.

\section{Models}
\label{sec:modelling}

\subsection{ISM size distribution}
\label{sec:ism_dist}
The ISM size distribution is usually described using a power law distribution:

\begin{equation}
    n (a) = A \; a^{\beta},   
    \label{eq:powerlaw}
\end{equation}
where $n$ is the number of particles per unit volume per unit size interval, $A$ is a normalization factor that depends on the assumed dust-to-gas ratio and the minimum and maximum grain sizes, and $\beta$ is the power law index that characterises how bottom- or top-heavy the size distribution is. The power law index $\beta$ and minimum and maximum grain sizes ($a_{min}$ and \amax) are typically chosen to be $-3.5$, $0.005 \mu$m, and $0.25 \mu$m, respectively, which fit the observed extinction law in the diffuse interstellar medium \citep{Mathis1977, Laor1993, Alexander1994a}. Although there are small variations in the values used for these parameters in the published literature, especially $a_{min}$ and \amax, they do not make an appreciable difference for the calculated opacity. The value for the normalizing constant $A$ is given by:

\begin{equation}
    A = \frac{3 \rho_{d} (\beta +4)}{4 \pi \rho_{s} (a_{\mathrm{max}}^{\beta+4} - a_{min}^{\beta+4})}, 
\end{equation}
where $\rho_s = 1.675$ g cm$^{-3}$ is the material density of the dust grain, fixed to a value appropriate for the DSHARP mixture (see \S~\ref{sec:dust_op_calc}), $\rho_d$ is the density of dust in the disk, $\rho_g$ the density of disk gas, and $\epsilon = \rho_d / \rho_g$ is the dust-to-gas ratio. The ISM dust opacity is typically calculated assuming a global value of $\epsilon = 0.01$ for the entire protoplanetary disk. For $\rho_g$, we use the gas density in the disk midplane obtained from our protoplanetary disk model, which we describe in the next section.

\subsection{Protoplanetary disk model}

We use the publicly available code \tpop to model the structure of a protoplanetary disk and the dynamics of dust and gas\footnote{The original public repository is available at \url{https://github.com/birnstiel/two-pop-py}. A fork of this repository with the changes implemented in our work is available at \url{https://github.com/y-chachan/two-pop-py/tree/rad_grad_d2g}.}. The methods and algorithms used in \tpop are described in \cite{Birnstiel2012} and we will present a brief overview here for completeness. We consider a protoplanetary disk of mass 0.1 M$_*$ around a protostar of mass $M_* = 0.7 \;$M$_{\odot}$. The stellar effective temperature ($T_*$) and radius ($R_*$) are set to $4010$ K and $1.806 \; R_{\odot}$ respectively. We assume that the disk is passively heated, and its temperature structure therefore takes the following form \citep[e.g.][]{Chiang1997, DAlessio1998}:

\begin{equation}
    T (r) = \bigg[ \phi \; T_*^4 \bigg( \frac{R_*}{r} \bigg)^2 + T_0^4 \bigg]^{1/4}
    \label{eq:temp_struct}
\end{equation}
where $r$ is the cylindrical distance from the star, $T_0 = 7$ K is a constant, and $\phi = 0.05$ is the angle between the incident radiation and disk surface (`flaring' angle). The sound speed $c_s$ is defined as $\sqrt{k_\mathrm{B} T / \mu m_H}$, where $k_\mathrm{B}$ is the Boltzmann constant, $\mu = 2.3$ is the mean molecular weight of the gas, and $m_H$ is the mass of hydrogen atom. Our neglect of heating due to viscous dissipation leads to an underestimation of the temperature in the inner regions of the disk but greatly simplifies the determination of the temperature structure. This choice does not have a significant effect on the position of the water ice line (see \S~\ref{sec:temp_struct} for a brief discussion). We note that accounting for the varying opacities that arise from the growth and transport of grains into the temperature profile is outside the scope of this paper \citep[see, e.g.][for recent attempts in this direction]{Savvidou2020}. 

The gas surface density ($\Sigma_{\mathrm{g}}$) is evolved following the fluid equations of viscously spreading accretion disk \citep{Lynden-Bell1974}

\begin{equation}
    \frac{\partial \Sigma_{\mathrm{g}}}{\partial t} = \frac{3}{r} \frac{\partial}{\partial r} \bigg[ r^{1/2} \frac{\partial}{\partial r} \big( \nu \Sigma_{\mathrm{g}} r^{1/2} \big) \bigg]
\end{equation}
whose self-similar solution (at time zero) is used to set the initial surface density profile for our calculation:

\begin{equation}
    \Sigma_{\mathrm{g}} (r) = C \bigg( \frac{r}{r_c} \bigg)^{-p} \mathrm{exp} \bigg[ -\bigg( \frac{r}{r_{\mathrm{c}}} \bigg)^{2-p} \bigg]
    \label{eq:init_gas}
\end{equation}
where $C$ is a constant to be normalized by the assumed disk gas mass, $\nu$ is the kinematic viscosity with a power law radial profile ($\nu = \nu_c(r/r_c)^p$), and $r_{\mathrm{c}}$ is a characteristic radius of the disk. Following \cite{Birnstiel2012}, we set $p = 1$ and $r_{\mathrm{c}} = 200$ au in our work. The viscosity $\nu = \alpha_{\mathrm{t}} c_s H_{\rm gas}$ is parameterized using the Shakura-Sunyaev turbulence parameter \alphat \citep{Shakura1973}, the sound speed $c_s$, and the gas scale height $H_{\rm gas} = c_s / \Omega$, where $\Omega$ is the Keplerian frequency.

The initial dust surface density is set as $\epsilon$ times the initial gas surface density given in Equation~\ref{eq:init_gas}, where $\epsilon = 0.01$. The dust surface density evolution and dynamics of dust is modelled using just two representative grain sizes in the disk (hence the name \tpop): the spatially and temporally constant monomer size $a_0$ and a large grain size $a_1$ that depends on time and location in the disk. We fix $a_0 = 0.005 \mu$m to align this variable with the minimum grain size in the ISM size distribution. These small grains rapidly coagulate to form agglomerates that are many orders of magnitude larger in size. Their growth is limited by processes such as turbulent fragmentation and radial drift. These limiting sizes are what set the value of $a_1$ as a function of time and $r$ and they are discussed in greater detail later in this section.

Splitting the dust population into two allows us to capture the qualitatively different dynamical behavior of large and small grains. Small grains are well coupled to the gas and are unable to maintain large relative velocities with respect to the gas. On the other hand, large grains are slightly decoupled from the gas and respond to pressure gradients on relatively short timescales. The total surface dust density ($\Sigma_{\mathrm{d}}$) is the sum of the surface density of small ($\Sigma_{0}$) and large ($\Sigma_{1}$) grains and can consequently be modelled using a single advection-diffusion equation:

\begin{equation}
    \frac{\partial \Sigma_{\mathrm{d}}}{\partial t} + \frac{1}{r} \frac{\partial}{\partial r} \bigg[r \bigg( \Sigma_{\mathrm{d}} \Bar{u} - D_{\mathrm{gas}} \Sigma_{\mathrm{g}} \frac{\partial}{\partial r} \bigg( \frac{\Sigma_{\mathrm{d}}}{\Sigma_{\mathrm{g}}} \bigg)  \bigg)  \bigg] = 0.
\end{equation}
Here, $\Bar{u}$ is the mass weighted radial velocity of dust grains and $D_{\mathrm{gas}}$ is the gas diffusivity. A derivation of this equation is available in the appendix of \cite{Birnstiel2012}.

The Stokes number St is dust grain stopping time under gas aerodynamic drag in units of local orbital time. Dust grains smaller than the gas particle mean free path are in Epstein drag regime and their Stokes numbers follow

\begin{equation}
    \mathrm{St} = \frac{\pi}{2} \frac{a \rho_s}{\Sigma_{\mathrm{g}}}.
    \label{eq:st_def}
\end{equation}
Detailed dust growth and evolution simulations indicate that grains will continue to grow until they reach a size (St $\sim 0.1 - 1$) where fragmentation due to collisions and/or loss to radial drift become significant \citep[e.g.][]{Brauer2008, Birnstiel2010}. For grains in this size range, velocity differences between grains due to turbulence become larger ($\Delta u \propto \sqrt{\mathrm{St}}$, \citealp{Ormel2007}) and collisions are more likely to lead to fragmentation instead of growth. This limits the maximum Stokes number and corresponding size $a_{\mathrm{frag}}$ that the grains can reach:

\begin{subequations}
\label{eq:afrag}
\begin{equation}
    \mathrm{St_{frag}} = \frac{1}{3 \alpha_{\mathrm{t}}} \frac{v_{\mathrm{frag}}^2}{c_s^2}
\end{equation}
\begin{equation}
    a_{\mathrm{frag}} = \frac{2}{3 \pi} \frac{\Sigma_{\mathrm{g}}}{\rho_s \alpha_{\mathrm{t}}} \frac{v_{\mathrm{frag}}^2}{c_s^2} 
\end{equation}
\end{subequations}
where \vfrag is the fragmentation velocity of dust grains.

The rate of radial drift is maximized for particles marginally coupled to gas (St $\sim$ 1) \citep{Weidenschilling1977, Chiang2010}:

\begin{equation}
    u_{\mathrm{drift}} = - \frac{2 u_{\eta}}{\mathrm{St} + \mathrm{St}^{-1}}
\end{equation}
where $u_{\eta} = - \gamma c_s^2 / 2 v_K$ is the drift velocity, $v_K$ is the Keplerian velocity, and $\gamma = |$d ln $P$ / d ln $r|$ is the power law index characterising the dependence of pressure on distance from the star. In some regions of the disk, particles may drift radially faster than they can grow to the size at which fragmentation dominates.  In these regions, the radial drift sets an upper limit on the particle size $a_{\mathrm{drift}}$:

\begin{equation}
    a_{\mathrm{drift}} = \frac{2}{\pi} \frac{\Sigma_{\mathrm{d}}}{\rho_s \gamma} \frac{v_K^2}{c_s^2}
    \label{eq:adrift}
\end{equation}
At early times in the disk evolution, the particle growth rate can also be a limiting factor for grain growth and set the maximum particle size. This can be true even at late times in the outer disk where the growth timescales ($\tau_{\mathrm{grow}} \simeq 1 / \epsilon \Omega$) are longer.
Relative velocities due to radial drift can also lead to fragmentation, but this effect is only relevant at early times for models with low turbulence ($\alpha_{\rm t} = 10^{-4}$). As the dust-to-gas ratio in such a region declines due to inward drift, the size limit set by radial drift becomes smaller than the one set by drift-induced fragmentation. In the two population model for dust evolution, the large grain size $a_1$ is fixed to a fraction of the maximum grain size that is determined by calibrating the \tpop model to the full simulations \citep{Birnstiel2012}. The maximum particle size limit therefore plays an important role in determining the dynamics of the large grains in the disk. Since most of the dust mass tends to be concentrated in the largest grains, which are also the most susceptible to radial drift, the dust-to-gas ratio of the disk can evolve significantly over time.

The turbulence parameter \alphat and the fragmentation velocity \vfrag are two of the most important parameters for determining the maximum particle size. The classically quoted range of values for \alphat is $10^{-4} - 10^{-2}$ \citep[e.g.][]{Turner2014}. However, recent studies of line broadening and dust settling in protoplanetary disks suggest that \alphat is closer to the lower end of this range \citep{Mulders2012, Pinte2016, Flaherty2015, Flaherty2017, Flaherty2018}. We therefore adopt $\alpha_{\mathrm{t}} = 10^{-3}$ for our baseline model and comment on the consequences of varying \alphat in \S~\ref{sec:alpha_vfrag_var}.\footnote{We note that we use the same $\alpha_t$ for both the global disk gas evolution and the turbulent stirring of dust. In reality, these two can be different \citep[see, e.g.,][]{Carrera2017,Drazkowska2017}.}

Both theoretical studies and experiments have long suggested a significant difference between the fragmentation velocities of ice-free and icy dust \citep{Poppe2000, Blum2008, Wada2013, Gundlach2015}. Most commonly, ice-free silicate dust is assumed to have a fragmentation velocity of 1 m/s, while icy grains have a fragmentation velocity closer to 10 m/s \citep[e.g.][]{Birnstiel2010, Pinilla2016, Drazkowska2017}. Such a difference in fragmentation velocity would lead to an abrupt change in the dust emission spectral index at water ice line \citep{Banzatti2015} and there is observational evidence to support the occurrence of this phenomenon \citep{Cieza2016}. This increase in fragmentation velocity for dust exterior to the water ice line has also been invoked to explain the architecture of the solar system and exoplanetary systems \citep[e.g.][]{Morbidelli2015, Venturini2020} as well as planetesimal formation \citep{Drazkowska2017}.

Despite this apparent consensus, recent theoretical and laboratory studies have begun to cast doubt on this story. Previous studies attributed the change in \vfrag to an order of magnitude difference in the surface energies of icy and ice-free dust grains, but recent experimental work now suggests that their surface energies may in fact be quite similar \citep{Gundlach2018, Steinpilz2019}. Other studies conclude that the fragmentation velocity might exhibit a more complicated and non-monotonic dependence on temperature \citep[e.g.][]{Gundlach2018, Musiolik2019}, and this topic remains an area of active debate in the community \citep[e.g.][]{Kimura2020a}. In this study we adopt the standard values of 1 m/s for ice-free and 10 m/s for icy grains for our baseline case, as these are close to the values derived from dynamical collision experiments.  We assume that the ice line is located where the disk temperature drops below approximately $T = 200$ K, which places the ice line at 0.75 au in all of our models. We use Gaussian convolution to smoothly increase \vfrag from 1 m/s at $250$ K to 10 m/s at $150$ K \citep[e.g.][]{Birnstiel2010}. In \S~\ref{sec:alpha_vfrag_var}, we also present alternative models where we vary the value of \vfrag both within and beyond the ice line and show that our results are qualitatively similar for a significant part of the plausible parameter space. 

We utilize the approximations from \cite{Birnstiel2015} (Equation 6, 7, and 8 in their paper) that are implemented in \texttt{twopoppy} to reconstruct the full grain size distribution in the protoplanetary disk, which we need in order to calculate the corresponding dust opacity. We also modified the \texttt{twopoppy} code to include the size limit set by drift-induced fragmentation in the size distribution calculation, which in the default version of the code is approximated by the radial drift-limited grain size instead (see \citealp{Birnstiel2012} for a discussion on the validity of this approximation). These approximations match the detailed simulations reasonably well, but can underestimate the number density of small grains. Although this will affect the opacity of the disk at short wavelengths (e.g., $\sim 1$ $\mu$m), we find that it only has a modest effect on the Rosseland mean opacity. We quantify this effect by comparing the mean opacity from this approximate method to the more accurate coagulation-fragmentation models from \cite{Birnstiel2011} in the fragmentation dominated region of the protoplanetary disk and find that the opacity from the approximate method is a factor of two smaller. In regions dominated by radial drift, a change in the assumed power law index for the size distribution can also affect the number of small particles. However, since radial drift tends to dominate in the outer colder parts of the disk, the mean opacity in this region is dominated by slightly larger grains ($\sim 100 \mu$m), which have a more robustly determined number density.

So far, we have discussed grain sizes, distributions, and opacities in the framework of a vertically integrated (2D) disk. If we wish to explore the 3D disk structure, we can extend these 2D models by using some reasonable approximations to calculate the density of dust and gas as a function of height from the midplane. This exercise is particularly valuable for planet formation models because growing protoplanets might not accrete most of their gas from the midplane (see \S~\ref{sec:planet_formation}). We assume a Gaussian vertical profile with a scale height $H_{\rm gas} (r) = c_s / \Omega$ for the gas. The midplane gas density is then given by $\rho_{\mathrm{g},0} = \Sigma_g / \sqrt{2 \pi} H_{\rm gas}$ (Equation~\ref{eq:st_def}, which gives the expression for St in the midplane, also used this assumption). Dust sediments towards the midplane and is carried upward by turbulent diffusion so its vertical density distribution is significantly different from that of the gas. We use the expression for the steady-state vertical distribution of dust derived by \cite{Fromang2009}:

\begin{equation}
    \rho_{\mathrm{d}} (z, a) = \rho_{\mathrm{d},0} \; \mathrm{exp} \bigg[ - \frac{\mathrm{St}_0}{\alpha_{\mathrm{t}}} \bigg(\mathrm{exp} \bigg( \frac{z^2}{ 2 H_{\rm gas}^2} \bigg) - 1 \bigg) - \frac{z^2}{ 2 H_{\rm gas}^2} \bigg]
    \label{eq:fromang}
\end{equation}
where $\rho_{\mathrm{d},0} (a)$ is the dust density and $\mathrm{St}_0 (a)$ is the Stokes number in the midplane for a particular grain size. In reality, the vertical scale height for dust should be set by either turbulent diffusion or the Kelvin-Helmholtz shear instability, depending on which is larger at a given disk location \citep{Rosenthal2018}. We find that for our fiducial model turbulent diffusion sets the dust scale height throughout the disk. The Kelvin-Helmholtz shear instability only comes into play for models with low turbulence strength ($\alpha_{\rm t} = 10^{-4}$) at large distances ($30-100$ au) and early times ($0.1 -1$ Myr). For this subset of models, incorporating its effect on the vertical dust distribution decreases the final gas-to-core mass ratio for a $15 \, M_{\oplus}$ core by at most 15\% if accretion stops at 1 Myr. Continued gas accretion beyond this time wipes out the effect of incorporating Kelvin-Helmholtz instability in our analysis. Since accounting for Kelvin-Helmholtz instability has a negligible impact on our results, we choose to omit it from our work.

\subsection{Calculation of dust opacity}
\label{sec:dust_op_calc}
The composition of dust grains in protoplanetary disks is a topic of active research \citep[see recent review by][]{Oberg2020}. We adopt the grain composition prescribed in the DSHARP survey papers and use the publicly available tools generously provided by the survey team for the calculation of grain properties \citep{Birnstiel2018}. The DSHARP composition mixture consists of water ice \citep[optical properties from][]{Warren2008}, `astrosilicates' \citep{Draine2003a}, and refractory organics and troilite (FeS) \citep{Henning1996}. The Bruggeman mixing rule is employed to obtain the optical constants for the mixture. We adopt the same grain composition for the entire disk, as removing water from our mixture has only a small effect ($\lesssim 15$\%, accounting for the difference in grain densities and optical properties but keeping the grain size distribution fixed) on the calculated opacity. Our simulations also do not account for the effect of condensation/sublimation on grain size and mass for particles moving across the ice line when calculating the grain size distribution. For the adopted DSHARP mixture,  water's sublimation would reduce dust mass only by 20\% within the ice line. Accounting for the reduced mass and increased density of ice free grains would reduce the grain size by $\sim 15$\%, which will have some effect on their dynamics. However, these effects are negligible compared to the other sources of uncertainty in our model.

We use Mie theory to calculate the dust opacity. Our Mie code is publicly available as part of \texttt{PLATON} \citep{Zhang2019, Zhang2020}, which uses the algorithm outlined by \cite{Kitzmann2018}. For particle sizes and wavelengths for which the full Mie treatment is impracticable, we resort to widely used approximations. We use the geometric optics limit to calculate the absorption cross-section of particles for which $|m| x > 1000$ and $|m-1| x > 0.001$, where $m$ is the complex refractive index of the particle and $x = 2 \pi a / \lambda$ is the size parameter (here $a$ being the particle size and $\lambda$ being the wavelength, \citealp{VandeHulst1957}). Specifically, we use the approximation described in \cite{Laor1993}, which uses the extinction coefficient calculated using Rayleigh-Gans approximation ($Q_{\mathrm{RG}}$) to obtain the extinction coefficient in the geometric optics limit ($Q_{\mathrm{ext}}$):

\begin{subequations}
\begin{equation}
    Q_{\mathrm{ext}} \approx \frac{Q_{\mathrm{RG}}}{(1 + 0.25 Q_{\mathrm{RG}}^2)^{1/2}}
\end{equation}
\begin{equation}
    Q_{\mathrm{RG}} = \frac{32 |m-1|^2 x^4}{27 + 16 x^2} + \frac{8}{3} \mathrm{Im} (m) x
\end{equation}
\end{subequations}
where Im$(m)$ is the imaginary part of the refractive index.

Once we have calculated the absorption coefficient for different particle sizes $a$ and wavelength $\lambda$, the wavelength dependent opacity $\kappa_{\lambda, a}$ for each particle size per gram of dust is given by:

\begin{equation}
    \kappa_{\lambda, a} = \frac{\pi a^2 Q_{\mathrm{ext}} (\lambda, a)}{4 \pi \rho_s a^3 / 3}
\end{equation}
To calculate the opacity per gram of dust in the protoplanetary disk, we need the normalized size distribution of the grains at a specific location. We utilize the mass density distribution of dust $\Sigma_{\mathrm{d}} (r, a)$, calculated in logarithmic bins of grain size using \tpop. The opacity per gram of dust in the protoplanetary disk is then obtained using:

\begin{equation}
    \kappa_{\lambda} = \frac{\int \kappa_{\lambda, a} \Sigma_{\mathrm{d}} (r, a) \; \mathrm{d \; ln} a}{\int \Sigma_{\mathrm{d}} (r, a) \; \mathrm{d \; ln} a}
\end{equation}
This wavelength dependent opacity is used to calculate the Rosseland mean opacity per gram of dust:

\begin{equation}
    \frac{1}{\kappa_{R}} = \frac{\int_{0}^{\infty} (1/\kappa_{\lambda}) (\partial B_{\lambda} / \partial T) d \lambda}{\int_{0}^{\infty} (\partial B_{\lambda} / \partial T) d \lambda}
\end{equation}
where $B_{\lambda}$ is the Planck function and $T$ is the temperature used in our protoplanetary disk model. To obtain the Rosseland mean opacity per gram of protoplanetary disk material, we multiply the $\kappa_{R}$ obtained above by the local dust-to-gas ratio $\epsilon = \Sigma_d (r) / \Sigma_g (r)$ of the disk. We do not include the gas opacity in our calculations, as the dust opacity dominates even in the regions with the lowest dust-to-gas ratio and/or the largest particle sizes (see \S~\ref{sec:gas_accretion}).

\section{Dust opacity in protoplanetary disks}
\label{sec:results}

\begin{figure}
    \centering
    \includegraphics[width=\linewidth]{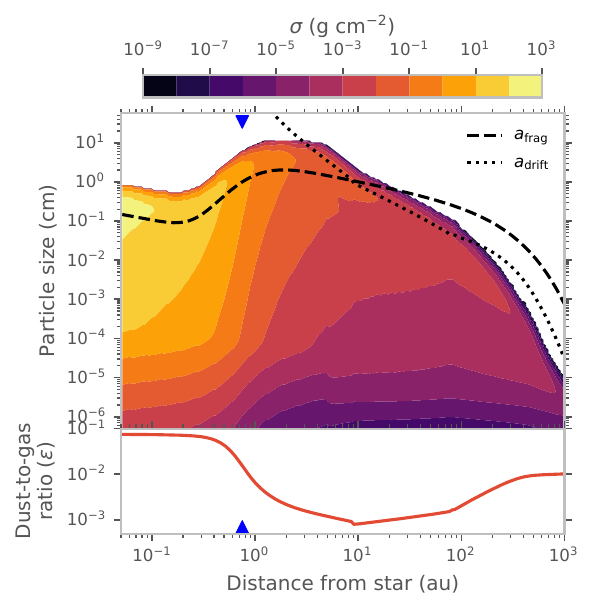}
    \caption{Size distribution and dust-to-gas ratio ($\epsilon$) at time $t = 1$ Myr for a \tpop simulation with variable \vfrag and $\alpha_{\mathrm{t}} = 10^{-3}$. The size limits imposed by fragmentation and drift are shown with dashed and dotted line in the upper panel. The location of the ice line is marked with blue triangles.}
    \label{fig:fiducial_size_dist}
\end{figure}

\subsection{Opacity from a simulated size distribution}
\begin{figure}
    \centering
    \includegraphics[width=\linewidth]{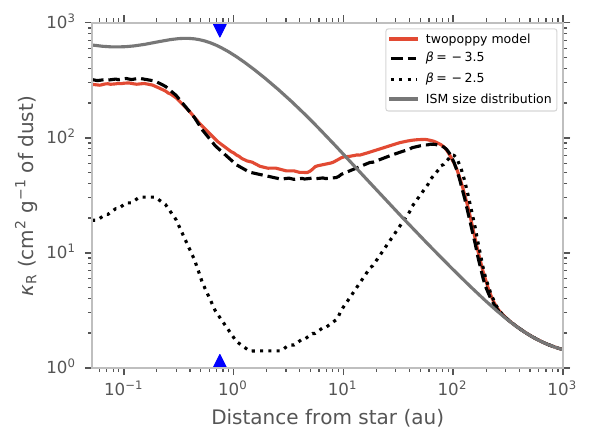}
    \caption{Rosseland mean opacity per gram of dust as a function of distance from the star at time $t = 1$ Myr. We adopt $\alpha_{\mathrm{t}} = 10^{-3}$ and a variable \vfrag that changes across the water ice line for our \tpop model. For the power law distributions, \amax is set by the location specific maximum grain size calculated from \tpop, which is given by Equation~\ref{eq:afrag} (fragmentation-limited), Equation~\ref{eq:adrift} (drift-limited), or the growth-timescale limit. The location of the ice line is marked with blue triangles.}
    \label{fig:sim_vs_pl_dust}
\end{figure}

In this section, we focus on quantifying the changes in the dust opacity due to location-dependent variations in the dust size distribution. We show the full radially-varying \tpop size distribution in the top panel of Figure~\ref{fig:fiducial_size_dist} and the resulting Rosseland mean opacity per gram of dust in Figure~\ref{fig:sim_vs_pl_dust}. The size distribution in the inner $10$ au is dominated by coagulation-fragmentation equilibrium, while the increase in \vfrag beyond the water ice line at $\sim 1$ au manifests as an increase in the maximum grain size ($a_{\mathrm{frag}} \propto v_{\mathrm{frag}}^2$ from Equation~\ref{eq:afrag}). Since larger grains contain more mass and the size distribution is slightly top-heavy, this increase in \vfrag causes the surface density of small grains ($\lesssim 10 \mu$m) to decrease by multiple orders of magnitude. Because these grains contribute significantly to \kr, this change is responsible for the factor of $\sim 5$ decrease in the simulated \kr shown in Figure~\ref{fig:sim_vs_pl_dust}. Beyond $\sim 10$ au, the maximum grain size is set by radial drift of the large grains instead of fragmentation as particles drift inward before they can grow to the fragmentation barrier. Without fragmentation to replenish the supply of small grains, the size distribution in this region becomes more top heavy relative to the distribution produced by the coagulation-fragmentation equilibrium in the inner disk. \kr in this cold outer disk region is dominated by larger grains ($\sim 100$ $\mu$m) that are relatively abundant, leading to a modest increase in the simulated \kr as shown in Figure~\ref{fig:sim_vs_pl_dust}.

In Figure~\ref{fig:sim_vs_pl_dust}, we compare \kr for size distribution simulated by \tpop at time $t = 1$ Myr with three different grain size distributions: the ISM size distribution ($\beta = -3.5, a_{\mathrm{max}} = 0.25 \mu$m) and power law distributions with $\beta$ of either $-2.5$ or $-3.5$ and maximum particle sizes set to the fragmentation (Eq.~\ref{eq:afrag}), radial drift (Eq.~\ref{eq:adrift}), or growth-timescale limits, as appropriate. We find that the dust opacity for the simulated size distribution differs significantly from that of the ISM size distribution (see also \citealp{Savvidou2020}). The opacity of the ISM size distribution only varies as a consequence of the decreasing temperature in the disk. In contrast, opacity from the simulated size distribution reflects radially varying grain growth and transport processes in the disk (e.g., \citealp{Akimkin2020} make the same observation). It is noteworthy that a power law distribution with $\beta = -3.5$ (same as that of the ISM) and \amax set by the relevant physics of fragmentation and radial drift yields a \kr profile that is in good agreement with the simulated results.

\begin{figure}
    \centering
    \includegraphics[width=\linewidth]{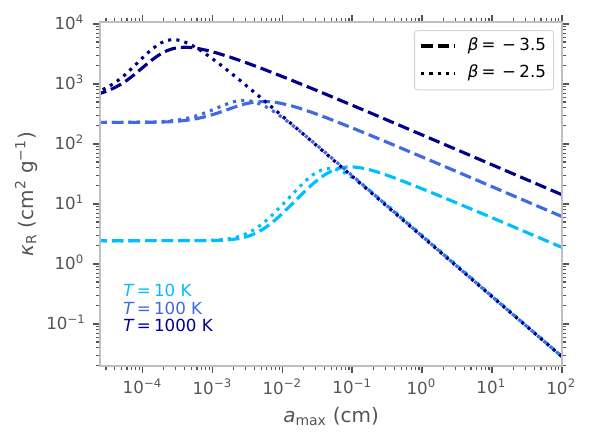}
    \caption{Rosseland mean opacity per gram of dust for a power law grain size distribution with $\beta = -3.5$ and $-2.5$ and three different temperatures. The lowest value of $a_{\mathrm{max}} = 0.25 \mu$m on this plot is the commonly adopted value for the ISM size distribution.}
    \label{fig:k_R_power_law_amax}
\end{figure}

We illustrate the effect of the maximum grain size \amax and the power law index $\beta$ on \kr in Figure~\ref{fig:k_R_power_law_amax}. The smallest value of \amax shown on the plot corresponds to the ISM size distribution. For top heavy distributions with $\beta > -4$, most of the mass is concentrated in the larger dust grains. Increasing \amax therefore redistributes dust mass from smaller grains to larger grains, reducing the total number of small grains. This can significantly alter the overall opacity of the dust: if we compare \kr for \amax = $0.1$ cm (which is more typical for dust in a disk) and $\beta = -3.5$ with the equivalent ISM value, it is almost 20 times larger at $10$ K. Conversely, this same depletion of smaller grains for \amax $= 0.1$ cm means that \kr is half the corresponding ISM value at $1000$ K. Using a realistic \amax for the power law size distribution of dust in a protoplanetary disk  therefore leads to a reduced \kr in the hotter inner disk and an enhanced \kr in the colder outer disk.

In contrast to this result, the opacity from a power law size distribution with $\beta = -3.5$ and \amax set by Equations~\ref{eq:afrag} and \ref{eq:adrift} and growth timescale $\tau_{\mathrm{grow}}$ provides a relatively good match to the opacity from the full simulated size distribution. The power law size distribution with $\beta = -2.5$ does not perform as well; this is due to the top heaviness of the $\beta = -2.5$ size distribution, which leads to a dramatic depletion in the number of small grains. Since the small grains that contribute most significantly to \kr at the protoplanetary disk temperatures are absent, the opacity for $\beta = -2.5$ is $\gtrsim 1$ order of magnitude lower than that for our \tpop simulation.  These results for different $\beta$ values are similar to previous findings for the dust opacity at specific wavelengths \citep[e.g.][]{DAlessio2001}.

\subsection{Opacity from a radially varying dust-to-gas ratio}

\begin{figure}
    \centering
    \includegraphics[width=\linewidth]{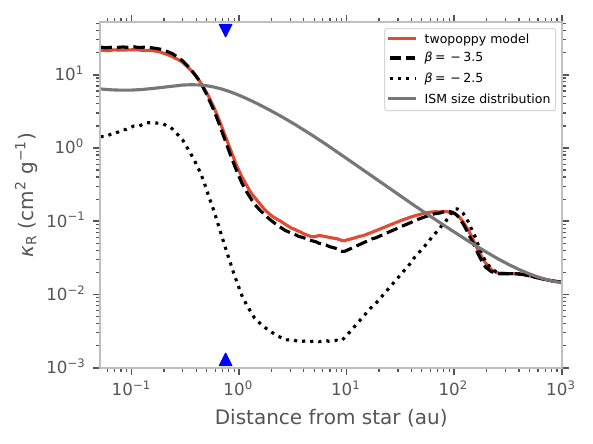}
    \caption{Rosseland mean opacity per gram of protoplanetary disk material at time $t = 1$ Myr and for a variable \vfrag and $\alpha_{\mathrm{t}} = 10^{-3}$. This plot is similar to Figure~\ref{fig:sim_vs_pl_dust} except that the opacity per gram of dust is here multiplied by the radially-varying dust-to-gas ratio. For the ISM size distribution, the dust-to-gas ratio is assumed to be 0.01 everywhere. The location of the ice line is marked with blue triangles.}
    \label{fig:sim_vs_pl_disk}
\end{figure}

Now that we have explored the effect of a radially varying dust size distribution on the Rosseland mean opacity per gram of dust, we can account for the radially varying dust-to-gas ratio $\epsilon$. As noted earlier, we assume that the contribution of the gas opacity to \kr is negligible. The dust-to-gas ratio (or metallicity) is typically fixed to a single global value \citep[e.g.][]{Bitsch2015, Mordasini2018}. However, this ratio can change radially as dust abundance evolves. Here we use our simulations to explore how the distribution of dust evolves in time as a function of assumed disk properties such as the turbulence strength \alphat and \vfrag.

\begin{figure*}
    \centering
    \includegraphics[width=\linewidth]{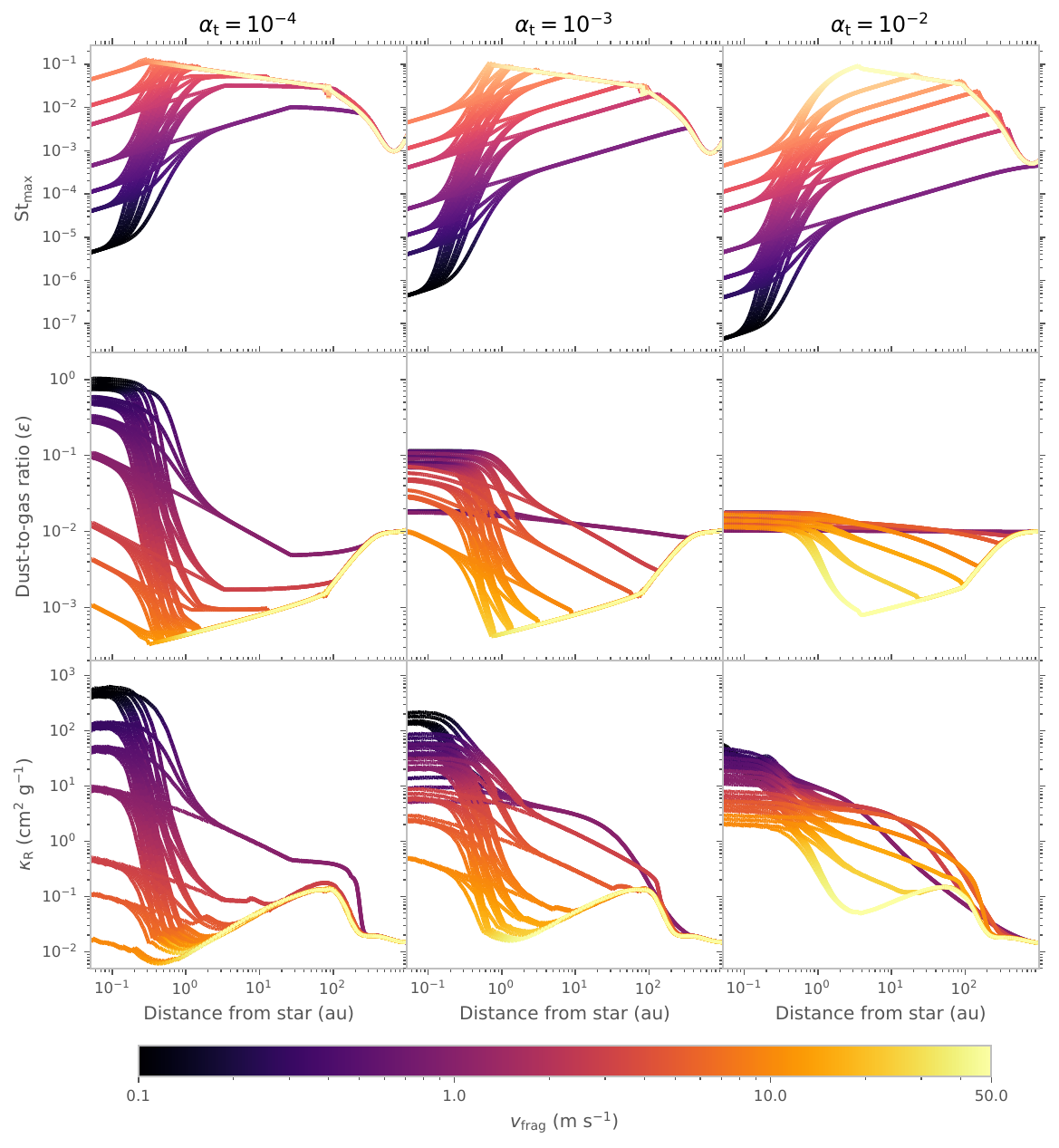}
    \caption{The Stokes number of the largest grain size (St$_{\mathrm{max}}$), dust-to-gas ratio ($\epsilon$), and Rosseland mean opacity per gram of protoplanetary disk material for a range of fragmentation velocities within and beyond the water snow line as well as three different turbulence strengths after 1 Myr of evolution. The fragmentation velocity \vfrag takes values in the range $0.1 - 10$ m s$^{-1}$ for ice-free grains and $1 - 50$ m s$^{-1}$ for icy grains \citep{Blum2008, Gundlach2015}. The ice line is located at 0.75 au in all of our models.}
    \label{fig:grid_plot}
\end{figure*}

We begin our simulation with a globally uniform $\epsilon = 0.01$ and show the resulting vertically integrated dust-to-gas ratio ($\epsilon = \Sigma_{\mathrm{d}} / \Sigma_{\mathrm{g}}$) at time $t = 1$ Myr for our fiducial model in the bottom panel of Figure~\ref{fig:fiducial_size_dist}. As grains begin to grow and their Stokes number increases, they face a stronger headwind from the gas and start drifting towards the star (see \citealp{Birnstiel2012} for a more detailed discussion on how the dust-to-gas ratio evolves in the disk). In the outermost regions of the disk ($\gtrsim 100$ au), the grain growth rate is so slow that particles do not reach the drift barrier, i.e. they do not drift very efficiently. $\epsilon$ far out does not evolve significantly and only decreases slowly as one moves closer to 100 au. Between $\sim 10$ and 100 au, grains drift inward faster than they can grow, causing the dust-to-gas ratio to decrease over time. In the inner disk, orbital timescales are shorter and grain growth is rapid.  This means that grains reach the fragmentation barrier before they can drift appreciably. For a fixed \vfrag and \alphat, the Stokes number of the largest grains also decreases as one moves closer to the star (see Eq.~\ref{eq:afrag}). This means that grains in the fragmentation-dominated inner disk are better coupled to the gas, and the dust-to-gas ratio does not decline as rapidly as in the drift-dominated outer disk region. In fact, the dust-to-gas ratio in the inner disk may even be enhanced by the migration of dust from the outer disk.

Depending on the magnitude of the velocity offset, the change in \vfrag across the ice line can have a dramatic effect on the dust dynamics. When large grains drifting inward from the outer disk cross the ice line they lose their ice and their fragmentation velocity decreases to the value characteristic of ice-free dust. Post-fragmentation grains are therefore smaller and their St is reduced, slowing their inward drift and causing a pile up of dust inside the ice line. The magnitude of this effect can be quite large: for a factor of $10$ decrease in \vfrag across the ice line, the St of the largest grains decreases by almost two orders of magnitude (St$_{\mathrm{frag}} \propto  v_{\mathrm{frag}}^2$). As shown in Figure~\ref{fig:fiducial_size_dist} this enhances the dust-to-gas ratio $\epsilon$ within $\sim 1$ au by almost an order of magnitude at $t = 1$ Myr relative to the starting $\epsilon$ of $0.01$. Conversely, most of the disk beyond $1$ au is significantly depleted of dust with $\epsilon \sim 10^{-3}$ for a large part of the outer disk. The effect of radial drift, fragmentation, and a change in \vfrag across the ice line on dust dynamics have been extensively described in \citet{Birnstiel2010, Pinilla2017}, and we refer the reader to these studies for a comprehensive exploration of this topic.

We can use this radially and temporally varying dust-to-gas ratio to update our calculation of the Rosseland mean opacity of the disk. Figure~\ref{fig:sim_vs_pl_disk} shows \kr per gram of protoplanetary disk material for our simulated size distribution. This plot is the same as Figure~\ref{fig:sim_vs_pl_dust} except that the \kr profiles shown in that figure are now multiplied by the dust-to-gas ratio. We plot the ISM \kr assuming a constant dust-to-gas ratio of $0.01$, in order to better illustrate the differences between our model and the widely used ISM opacity model. Within the ice line, the dust-to-gas ratio is enhanced by a factor of ten relative to the ISM model, which partially compensates for the reduction in opacity due to the increased grain sizes (Figure~\ref{fig:sim_vs_pl_dust}). As we move beyond the ice line, the decreasing quantity of dust and increasing concentration of dust mass in larger particle sizes lead to a steep decline in the opacity. Our \kr between $\sim$1 and $\sim$10 au is smaller than the ISM value by more than a factor of ten.

\subsection{Dependence on the assumed fragmentation velocity and turbulence strength}
\label{sec:alpha_vfrag_var}

Our fiducial model predicts that the dust opacity will decrease by more than two orders of magnitude as we move outside the ice line. However, the magnitude of this gradient depends strongly on the absolute and relative efficiency of dust transport in the inner and outer disk. The transition from the fragmentation-dominated to the drift-dominated regime can be expressed as a function of the fragmentation velocity $v_{\mathrm{frag}}$ and the turbulence strength \alphat \citep[e.g.][]{Birnstiel2015}:

\begin{equation}
    \frac{v_{\mathrm{frag}}^2}{v_{\mathrm{K}}^2} > \frac{3 \alpha_{\mathrm{t}} \epsilon}{\gamma}
    \label{eq:frag_to_drift}
\end{equation}
This transition also depends on the Keplerian velocity $v_{\mathrm{K}}$, the dust-to-gas ratio $\epsilon$, and $\gamma = |\mathrm{d ln} P / \mathrm{d ln} r|$. All of these quantities can vary as a function of $r$ (although we assume \alphat is constant in our work) and in regions where this inequality is satisfied, the disk becomes drift-dominated. Since \alphat and \vfrag are not known a priori, we run a grid of models over $\alpha_{\mathrm{t}} \in [10^{-4},10^{-3},10^{-2}]$ where $\alpha_{\mathrm{t}} = 10^{-3}$ is our fiducial, and $v_{\rm frag} = 0.1-10$ m s$^{-1}$ for ice-free grains and $1-50$ m s$^{-1}$ for icy grains \citep[e.g.][]{Blum2008, Gundlach2015}. We consider all possible combinations of these two fragmentation velocities as long as they meet the requirement that \vfrag for icy grains is greater than or equal to \vfrag for ice-free grains \footnote{We note that the \tpop models are calibrated with the full numerical models for a smaller range of \vfrag ($1 - 10$ m/s) than we study here. However, this should not be a major concern as the underlying collisional outcome model is the same.}.

\begin{figure*}
    \centering
    \includegraphics[width=\linewidth]{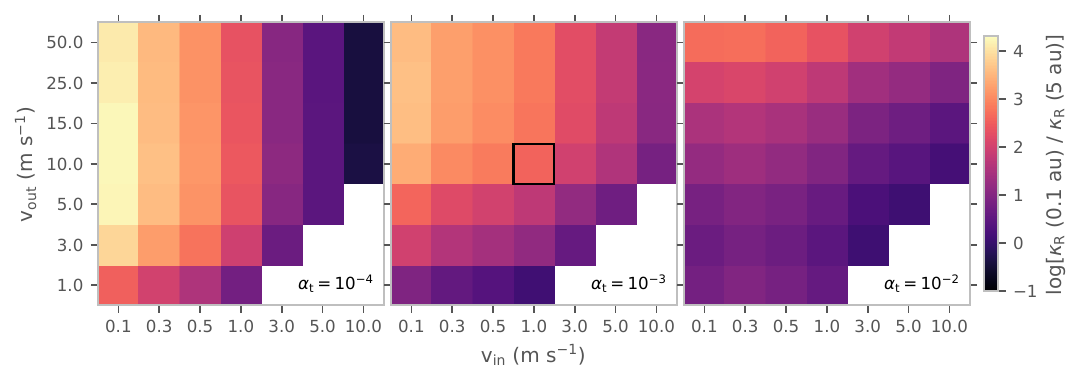}
    \caption{Ratio of the Rosseland mean opacity per gram of protoplanetary disk material at 0.1 au and 5 au after 1 Myr of evolution. The axes labels $v_{\mathrm{in}}$ and $v_{\mathrm{out}}$ stand for the fragmentation velicity within and beyond the ice line. Our fiducial model is outlined with a black square.}
    \label{fig:k_R_alpha_vfrag_grid}
\end{figure*}

Figure~\ref{fig:grid_plot} shows the Stokes number of the largest grains St$_{\mathrm{max}}$, the dust-to-gas ratio $\epsilon$, and the disk's Rosseland mean opacity \kr for this grid of models. As \alphat decreases, we find that $\epsilon$ varies more strongly with orbital distance: a consequence of the difference in the absolute values of St$_{\mathrm{max}}$ for different \alphat. For $\alpha_{\mathrm{t}} = 10^{-4}$, St$_{\mathrm{max}} \gtrsim 10^{-2}$ between $\sim$1 and 100 au. A larger Stokes number beyond the ice line leads to more efficient inward drift of dust from the outer to the inner disk. For lower \alphat, the transition to the drift-dominated region also happens closer in to the star (see Equation~\ref{eq:frag_to_drift} above), creating a `kink' in St$_{\rm max}$ and $\epsilon$ profiles (e.g. at 10 au in our fiducial model, see bottom panel of Figure~\ref{fig:fiducial_size_dist}). In the outer disk, all models that transition to the drift-dominated regime converge to similar values for St$_{\mathrm{max}}$ and $\epsilon$. For $\alpha_{\mathrm{t}} = 10^{-2}$ this transition moves outside $\sim 100$ au for most models, causing the disk to be globally fragmentation-dominated. As a result, St$_{\mathrm{max}}$ has a lower value throughout the disk and dust migration is suppressed.

The high St in the low \alphat disk model, which aids the radial transport of dust grains in the outer disk, can also potentially diminish the dust pile-up in the inner disk. Since St$_{\mathrm{frag}} \propto \alpha_{\mathrm{t}}^{-1}$, St$_{\mathrm{max}}$ in the inner disk is larger for lower \alphat. As long as St$_{\mathrm{max}} < 1$, the inward drift velocity will be larger for a larger value of St$_{\mathrm{max}}$.  This means that dust grains in the inner disk will move inward faster when \alphat is lower, reducing the timescale over which dust is depleted in the inner disk and preventing a pile-up of dust drifting in from the outer disk. This is evident in the middle panel of Figure~\ref{fig:grid_plot}, which shows that when $\alpha_{\rm t} = 10^{-4}$ the dust-to-gas ratio in the inner disk can be either very high ($\sim 1$, efficient pile-up, low \vfrag) or very low ($\sim 10^{-3}$, no pile-up, high \vfrag) depending on the assumed fragmentation velocities. Maximizing the dust-to-gas ratio and consequently the opacity gradient in the radial direction therefore requires an intermediate value of \alphat, which in turn is dependent on \vfrag.

Larger differences in the \vfrag values for icy and ice-free grains lead to a larger change in St$_{\mathrm{max}}$ across the ice line. This in turns results in a depletion of dust in the outer disk and a pile up of dust in the inner disk, leading to larger opacity contrast between the inner and the outer disk (see Figure~\ref{fig:k_R_alpha_vfrag_grid}), as long as the value of \alphat does not nullify these effects by either producing globally low values of St$_{\mathrm{max}}$ (well coupled dust and little dust transport) or large values of St$_{\mathrm{max}}$ within the ice line (dust drifts towards the star and does not pile up). However, when \vfrag is large everywhere in the disk (e.g. 10 m/s for ice-free grains and 50 m/s for icy grains), particles will have large St$_{\mathrm{max}}$ and will rapidly drain onto the star.

To simplify comparisons between models, in Figure~\ref{fig:k_R_alpha_vfrag_grid} we focus on the ratio of the disk opacity \kr at $0.1$ au and $5$ au. These distances are chosen to best capture the opacity contrast for the full set of disk models; they are also approximately where sub-Neptunes and gas giants are most numerous, respectively. We find that there is a large range of choices for \vfrag and \alphat that lead to opacity contrasts that are equal to or larger than the one in our fiducial model. Decreasing \alphat enlarges St$_{\mathrm{max}}$ and accelerates the grain radial transport, enhancing the contrast in the opacity across the snow line. Larger differences in \vfrag between icy and ice-free grains also produce greater opacity contrasts as they lead to a strong gradient in dust transport efficiency across the snow line.

The opacity contrast with increasing \vfrag for icy grains saturates at a value that depends on the \vfrag for ice-free grains. This is most evident in the lower \alphat models and occurs because St$_{\mathrm{max}}$ and $\epsilon$ converge to similar values in the outer disk (Figure~\ref{fig:grid_plot}). Beyond this limit, increasing the \vfrag for icy grains does not lead to an increase in the Stokes number of the largest grains in the outer disk but instead simply pushes the transition from fragmentation-dominated regime to drift-dominated regime inward. This limits the supply of dust from the outer disk and causes the opacity contrast to saturate at a fixed \vfrag for ice-free grains.

Recent observations of protoplanetary disks appear to favor values for \alphat that are lower than $10^{-2}$ \citep[e.g.][]{Pinte2016, Flaherty2018}. As we discussed earlier, it is less clear how large the difference in \vfrag for icy and ice-free dust grains may be \citep{Gundlach2018, Steinpilz2019, Kimura2020a}. However, our parameter space exploration suggests that there are a wide range of plausible scenarios that can lead to a large opacity gradient between the inner and outer disk regions.

\subsection{Dust opacity in a 3D disk}

So far, we have only considered vertically integrated disk models. In this section we examine the vertical structure of the dust distribution and its potential importance for planet formation (e.g., polar accretion of gas onto planetary cores; \citealt{Ormel2015, Fung2015, Cimerman2017, Lambrechts2017}) and modelling protoplanetary disks. The vertical structure of gas and dust is controlled by a complicated coupling between the disk temperature, opacity, and turbulence. Self-consistently taking these couplings into account is beyond the scope of our study; instead, we utilize a simple vertically isothermal disk model. Even with this simplification, our model produces a non-uniform vertical distribution of dust grains.

\begin{figure}
    \centering
    \includegraphics[width=\linewidth]{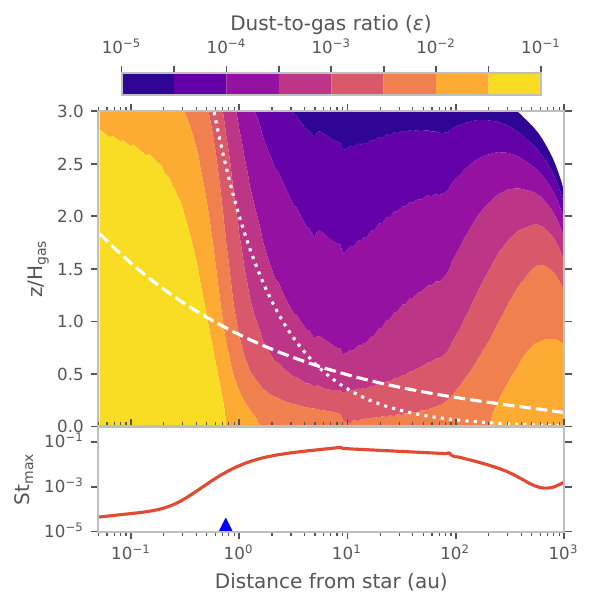}
    \caption{The top panel shows the dust-to-gas ratio $\epsilon$ as a function of height above the midplane $z$ and distance from the star after 1 Myr of evolution. The white dashed and dotted lines mark the height of the Hill radius $R_{\mathrm{Hill}}$ and Bondi radius $R_{\mathrm{Bondi}}$ of a 15 M$_\oplus$ planet respectively. The bottom panel shows the midplane Stokes number of the largest grains present in the disk at $t = 1$ Myr. The water ice line is marked with a blue triangle. Well coupled grains within the ice line lead to efficient vertical mixing of grains and hence a weak dependence of $\epsilon$ on $z$. Beyond the ice line, large grains that dominate the dust mass settle close to the midplane, which leads to a strong decline in $\epsilon$ as a function of $z$.}
    \label{fig:3d_d2g}
\end{figure}

\begin{figure}
    \centering
    \includegraphics[width=\linewidth]{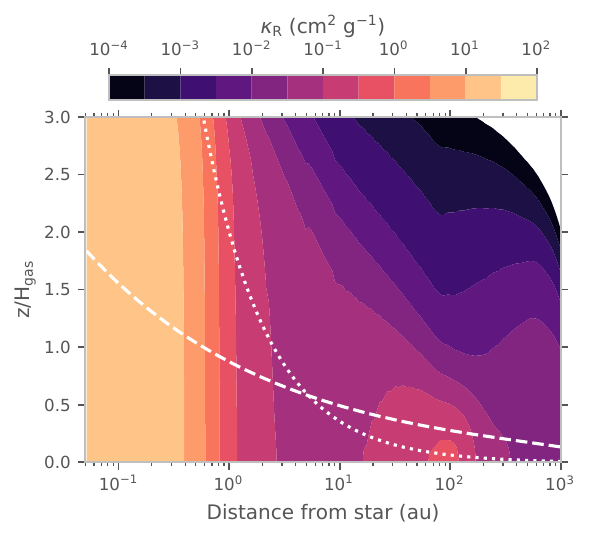}
    \caption{Rosseland mean opacity per gram of protoplanetary disk material as a function of height above the midplane $z$ and distance from the star after 1 Myr of evolution. The white dashed and dotted lines mark the height of the Hill radius $R_{\mathrm{Hill}}$ and Bondi radius $R_{\mathrm{Bondi}}$ of a 15 M$_\oplus$ planet respectively. Vertically well mixed dust within the ice line leads to little variation in \kr as a function of $z$. Grain settling and a strong decline in $\epsilon$ with $z$  leads to a gradient in \kr as a function of $z$ beyond the ice line.}
    \label{fig:3d_opacity}
\end{figure}

We use the prescribed radial temperature structure from Equation~\ref{eq:temp_struct} and assume a vertically isothermal disk structure in order to calculate the vertical structure of the dust and gas. Under this assumption, the gas density $\rho_g \propto e^{-z^2/H_{\rm gas}^2}$ where $z$ is the height from the midplane and $H_{\rm gas} = c_s / \Omega$ is the gas disk scale height. For the vertical dust density distribution, we utilize the expression obtained by \cite{Fromang2009} for the steady-state distribution of dust (Equation~\ref{eq:fromang}). We calculate the 3D dust density $\rho_d (z, a)$ for logarithmically binned grain sizes and sum it to obtain the total dust density $\rho_d (z)$. The dust-to-gas ratio $\epsilon$ is then simply calculated as $\rho_d / \rho_g$. 

The top panel in Figure~\ref{fig:3d_d2g} shows the resulting dust-to-gas ratio $\epsilon$ as a function of $z$ and distance from the star for our fiducial model at a disk age of 1 Myr. The differences in $\epsilon$ as a function of $z$ within and beyond the ice line can be understood by examining the Stokes number of the largest grains St$_{\mathrm{max}}$ present in each region of the disk (bottom panel of Figure~\ref{fig:3d_d2g}). Within $\sim$1 au, St$_{\rm max}$ can fall down to $\sim$10$^{-4}$; these particles will be vertically well-mixed with the gas---i.e. the scale height of dust grains is comparable to that of the gas---flattening the vertical gradient in dust-to-gas ratio. However, outside the ice line, large grains with St$_{\mathrm{max}} \gtrsim 10^{-2}$ are present. These grains are concentrated near the midplane and constitute most of the dust mass budget, resulting in a steep vertical gradient in $\epsilon$. Figure~\ref{fig:3d_opacity} shows the Rosseland mean opacity of the disk as a function of height from the midplane and distance from the star. As expected, we find that the disk opacity is essentially independent of $z$ within the ice line. In contrast, the concentration of large grains near the midplane beyond the ice line leads to a decline in disk opacity as a function of $z$. 

We mark the Hill radius $R_{\rm Hill} = a (M_{\rm p} / 3 M_*)^{1/3}$ and Bondi radius $R_{\rm Bondi} = G M_{\rm p} / c_s^2$ of a 15 M$_{\oplus}$ core with a dashed and dotted line respectively in Figures~\ref{fig:3d_d2g} and \ref{fig:3d_opacity}. We choose a mass of 15 M$_{\oplus}$ as our fiducial case as it is representative of a giant planet core. Planetary cores close to thermal or superthermal mass (equivalently, $R_{\rm Hill} \leq R_{\rm Bondi}$) are expected to accrete gas from heights on the order of the Hill radius \citep[e.g.][]{Lambrechts2017}. For subthermal cores (equivalently, $R_{\rm Hill} > R_{\rm Bondi}$), on the other hand, the natural length scale is expected to be the Bondi radius (see, e.g., subthermal cases of \citealt{Ormel2015} and \citealt{Fung2019}). The exact origin height of the accretion flow is unclear given how unsteady the flow morphology is in three-dimensional calculations. In this work, we assume that the material accreted by the planet is well represented by the properties of dust and gas present at min($R_{\rm Hill}, R_{\rm Bondi}$) above the disk midplane. In \S~\ref{sec:gas_accretion}, we show the effect of varying this height on the calculated gas-to-core mass fraction of a planet.

\begin{figure}
    \centering
    \includegraphics[width=\linewidth]{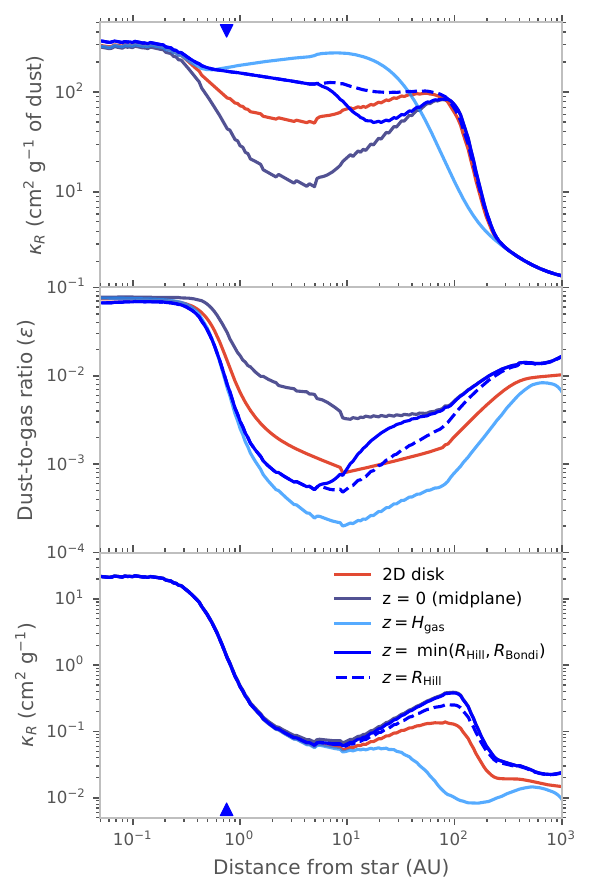}
    \caption{A comparison of the Rossleand mean opacity per gram of dust, dust-to-gas ratio $\epsilon$, and Rosseland mean opacity per gram of protoplanetary disk material \kr for our fiducial 2D disk integrated model and our 3D disk model after 1 Myr of evolution. We plot the values of these quantities in the disk midplane ($z = 0$), a single gas scale height above the midplane ($z = H_{\rm gas}$), and at heights of a 15 M$_{\oplus}$ planet's $R_{\mathrm{Hill}}$ and min($R_{\mathrm{Hill}}, R_{\mathrm{Bondi}}$) above the midplane. The water ice line is marked with blue triangles.}
    \label{fig:3d_z_comparison}
\end{figure}

Figure~\ref{fig:3d_z_comparison} highlights how the radial profile of dust-to-gas ratio and dust opacity differ for different heights above the disk midplane: $z = 0$ (disk midplane), $z = H_{\rm gas}$, and $z = R_{\rm Hill}$ and $z = \mathrm{min}(R_{\rm Hill}, R_{\rm Bondi})$ for a 15 M$_{\oplus}$ core. We also provide a calculation of the vertically integrated disk model for comparison. In the top panel we plot \kr per gram of dust, which depends only on the local size distribution of the dust. The features present in the \kr profiles result from changes in the relative abundances of the grain sizes that contribute most to the opacity at the local temperature. In the disk midplane beyond the ice line, most of the opacity contribution comes from grains that are $10 - 100 \, \mu$m in size but most of the mass (per gram of dust) resides in grains that are larger than this size range. This leads to a substantial decrease in \kr per gram of dust in the disk midplane in these regions. Conversely, the high relative abundance of small grains at $z = H_{\rm gas}$ (only small grains can be lifted to this height) leads to a strong enhancement in \kr per gram of dust at this height. The \kr profile at $z = R_{\mathrm{Hill}}$ and $z = \mathrm{min}(R_{\mathrm{Hill}}, R_{\mathrm{Bondi}})$ in the top panel of Figure~\ref{fig:3d_z_comparison} can be understood using these same principles.

Dust-to-gas ratio increases with higher concentration of large grains for a top heavy size distribution and so we observe a flipped behavior for the $\epsilon$ ratio profile (middle panel of Figure~\ref{fig:3d_z_comparison}) where it reaches lower values at higher altitudes beyond the ice line. Since larger grains settle close to the midplane, $\epsilon$ is highest at the midplane and decreases higher up. The $\epsilon$ evaluated at min($R_{\rm Hill}$, $R_{\rm Bondi}$) converges to that of the midplane in the innermost and the outermost region. The former arises from efficient vertical mixing whereas the latter materializes from $R_{\rm Hill}/H_{\rm gas}$ and $R_{\rm Bondi}/H_{\rm gas}$ approaching zero in the outer disk (see $R_{\rm Hill}$ and $R_{\rm Bondi}$ profiles in Figure~\ref{fig:3d_d2g}).

In the bottom panel of Figure~\ref{fig:3d_z_comparison} we plot the mean opacity per gram of protoplanetary disk material, which is the product of the quantities plotted in the upper two panels. Regardless of our vertical location in the disk, we see the same precipitous decline in disk opacity as in the vertically integrated disk model. Notably, \kr decreases by $\sim 2$ orders of magnitude between 0.1 au and 5 au at the height of our fiducial planetary core's $R_{\mathrm{Hill}}$. Within $\sim 10$ au, the \kr profiles for the vertically integrated disk model and the different $z$ values are nearly identical. This happens within the ice line as a result of efficient vertical mixing of grains (i.e. both \kr per gram of dust and $\epsilon$ are roughly constant as a function of $z$). Beyond the ice line and within 10 au, the sharp decline in $\epsilon$ with $z$ is counterbalanced by the increase in \kr per gram of dust with $z$ to yield a weakly $z$ dependent \kr (per gram of protoplanetary disk material).

\begin{figure*}
    \centering
    \includegraphics[width=0.49\linewidth]{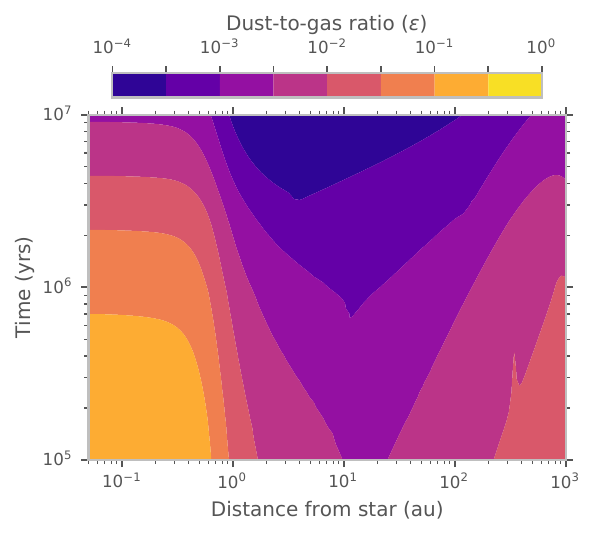}
    \includegraphics[width=0.49\linewidth]{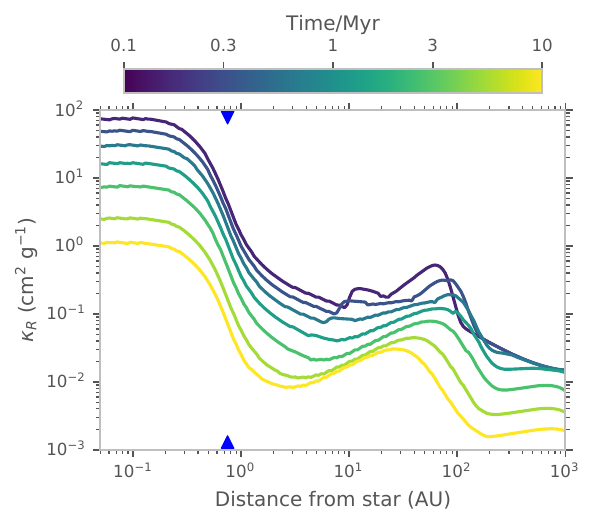}
    \caption{Time evolution of the vertically integrated dust-to-gas ratio $\epsilon$ and the Rosseland mean opacity per gram of protoplanetary disk material as a function of distance from the star. Although the absolute values of $\epsilon$ and \kr decline over time due to global accretion of dust onto the star, there is little change in their observed profile shapes as a function of time. The minima in $\epsilon$ and \kr profiles move slightly inward with time as a larger fraction of the outer disk becomes drift dominated. The water ice line is marked with blue triangles in the right panel.}
    \label{fig:op_time_ev}
\end{figure*}

\subsection{Time evolution of the dust opacity}

Up to this point we have presented results from our models after 1 Myr of disk evolution. In this section we explore the time-varying grain size distribution and dust-to-gas-ratio from 0.1 to 10 Myrs, where the lower limit is chosen to represent the plausible time at which massive planetary cores emerge. Figure \ref{fig:op_time_ev} demonstrates that the absolute values of the dust-to-gas ratio and mean opacity throughout the disk tend to decline over time. This is due to the global depletion of dust in the disk as it gradually accretes onto the star. Because the timescale over which $\epsilon$ and \kr evolve lengthens as time goes on, we present our results as a function of log time. Already by $0.1$ Myr, the dust-to-gas ratio and \kr profiles converge to shapes that are qualitatively similar to those of our fiducial 1 Myr model. Although temporal evolution of the disk after 0.1 Myr leads to $1 - 2$ orders of magnitude decline in the dust-to-gas ratio and opacity, it has a small effect on their radial gradient in the disk. However, there is a noticeable inward movement of the minima in $\epsilon$ and \kr profiles with time. This is because as the dust-to-gas ratio declines in the outer disk, the radius at which the disk transitions from being fragmentation-dominated to drift-dominated moves inwards (Equation~\ref{eq:frag_to_drift}). As we will show in \S~\ref{sec:planet_formation}, the overall decline in $\epsilon$ and \kr over time leads to the enhancement of gas accretion onto planetary cores.

We note that the assumed disk size also plays an important role in the temporal evolution of the dust-to-gas ratio and consequently dust opacity. Due to radial drift, dust drains onto the star more rapidly in smaller disks and the dust-to-gas ratio in the outer disk can become very small ($\lesssim 10^{-4}$) as early as 1 Myr. Radial drift may be too efficient in disk models and there is some tension with observations \citep[e.g.][]{Takeuchi2005, Brauer2008}, as many disks with ages of a few Myr appear to have mm-sized grains present at large distances ($\gtrsim 100$ of au) \citep[e.g.][]{Andrews2018, Hendler2020}. Proposed solutions for resolving the radial drift problem include the presence of dust traps \citep[e.g.][]{Kretke2007, Pinilla2012, Zhu2014}, larger than assumed disk gas density \citep{Powell2019}, and grains with large porosity \citep{Estrada2015, Garcia2020}. We circumvent this issue by modeling a relatively large and massive disk, ensuring a reasonable supply of dust throughout the disk lifetime.

\subsection{Temperature structure of the disk}
\label{sec:temp_struct}

\begin{figure}
    \centering
    \includegraphics[width=\linewidth]{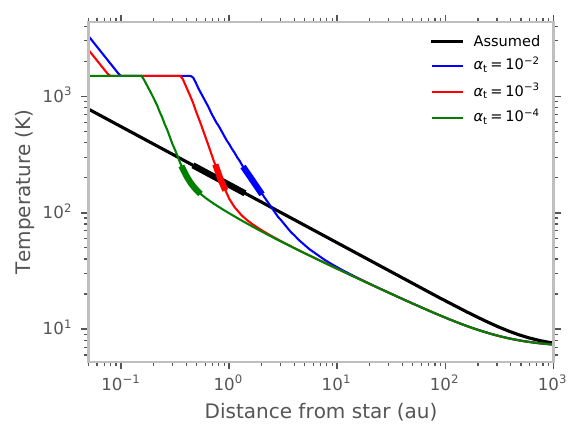}
    \caption{Post-processed temperature structure of the disk calculated using the method outlined in \cite{Birnstiel2010} with a self-consistent treatment of the opacity calculated using the size distribution and dust-to-gas ratio from our simulations. The disk properties at $t = 1$ Myr are used to calculate the temperature structure and are taken from simulations for which \vfrag is 1 m/s and 10 m/s for ice-free and icy grains respectively. The water snow line ($150-250$ K) is marked with thick lines for each temperature profile.}
    \label{fig:post_proc_T_struct}
\end{figure}

We have calculated the disk's temperature profile assuming a passively irradiated disk and thus assumed that it is identical for all of our grid models. However, accretional heating plays a role in setting the temperature structure of the disk, especially in the inner region and in particular for high disk viscosity (high \alphat). In addition, dust dynamics and a variable \vfrag alter the dust-to-gas ratio and the dust size distribution through the disk, leading to a location dependent dust opacity. Ideally, our models would include a self-consistent coupling of the dust and gas dynamics with the disk temperature structure including the effect of heating due to accretion. Although this is beyond the scope of this work, we carry out a preliminary assessment of the effect that accretional heating and enhanced dust-to-gas ratio would have on the temperature profile. In particular, we are interested in whether accretional heating has a significant impact on the location of the water ice line, which marks the transition in \vfrag values.

We post-processed results from our simulations to calculate the temperature structure of the disk using the method outlined in \cite{Birnstiel2010}. The only difference in our method is that we self-consistently calculate the Rosseland and Planck mean opacities using the size distribution and dust-to-gas ratio from our simulations. To speed up the calculation of dust opacity, we use a power law size distribution for the dust with the power law index $\beta = -3.5$ and \amax set by our \tpop simulations (see \S~3.2). We also account for the thermostat effect of dust’s vaporisation at high temperatures \citep{Birnstiel2010}. When the temperature reaches 1500 K, any further increase in the temperature leads to the vaporisation of dust (and a decrease in opacity) so temperature is kept stabilized at this value. Once the gas opacity alone (assumed to be 0.1 cm$^2$ g$^{-1}$) is enough to raise the temperature above 1500 K, we assume that all of the dust has evaporated and allow the temperature to rise again.

We find that accounting for the effects of accretional heating and the elevated dust-to-gas ratio significantly increases the temperature of the disk inside $\sim$1 au, but it has a negligible impact on the location of the water snow line (Figure~\ref{fig:post_proc_T_struct}). For our fiducial model with \alphat $= 10^{-3}$, the snow line location is essentially identical. For \alphat $= 10^{-4}$ and $10^{-2}$, the snow line moves from 0.75 au to 0.4 au and 1.6 au, respectively. We speculate that the increased temperature in the inner disk could also have an impact on the dust dynamics as it would lead to a decrease in St$_{\rm frag}$ ($\propto 1/c_{\rm s}^2$). This might further reduce the rate at which dust in the inner disk drains onto the star, thereby resulting in a larger dust pile up.  This reinforces our conclusions about the difficulty of accreting gas in this region. However, obtaining a full solution to this problem would require us to couple the dust and gas dynamics with the disk temperature structure, which to the best of our knowledge has only been attempted once in the published literature \citep{Estrada2016}.

\section{Implications for planet formation}
\label{sec:planet_formation}

\subsection{Gas accretion mediated by cooling}
\label{sec:gas_accretion}

Our calculated values for the dust opacity as a function of distance from the star show a dramatic decrease as we move beyond the ice line. We now consider what effect this variation in dust opacity and dust-to-gas ratio might have on the ability of planetary cores to accrete hydrogen-rich envelopes. For cores with masses $\lesssim 20$ M$_\oplus$, the rate of gas accretion onto the planetary core is initially regulated by the envelope's ability to cool and contract \citep[e.g.,][]{Lee2019}. This cooling is controlled by the properties of the gas envelope at the innermost radiative-convective boundary (RCB), as most of the cooling luminosity is generated inside the innermost convective zone \citep{Piso2014, Lee2014}.

There is a qualitative difference in the radiative-convective structure of planetary envelopes dominated by dust opacity versus gas opacity. For `dust-free' envelopes with negligible dust opacity, we expect to see a single convective zone that is connected to the disk via a nearly isothermal radiative zone. However, for `dusty' envelopes where dust opacity dominates over gas opacity, the evaporation of dust grains deep inside the envelope leads to a dramatic drop in the local envelope opacity, which causes an intermediate radiative zone to form. \cite{Lee2014} show that in this case, the innermost RCB appears at the H$_2$ dissociation front ($\sim 2500$ K) where H$^-$ opacity starts to dominate.

We expect atmospheres to transition to the `dust-free' accretion regime when the dust opacity is comparable to the gas opacity at the relevant temperature. In the inner disk ($\sim 0.1$ au), this transition occurs when the dust opacity approaches $\sim 0.01$ cm$^2$ g$^{-1}$. As we move farther out in the disk, the gas opacity decreases sharply ($\lesssim 10^{-4}$ cm$^2$ g$^{-1}$ at the relevant densities; e.g. \citealp{Freedman2014}) as the number of available molecular line transitions decreases. In our fiducial disk model for a 15 M$_\oplus$ core, the dust opacity at a height of min($R_{\mathrm{Hill}}$, $R_{\mathrm{Bondi}}$) above the midplane does not go below the gas opacity limit. Our models therefore predict that accretion at all orbital distance occurs in the `dusty' regime, whose RCB opacity---which controls the rate of cooling and therefore accretion---is given by the H$^-$ opacity \citep{Lee2015}:

\begin{multline}
    \kappa (\mathrm{H}^-) \sim 3 \times 10^{-2} \mathrm{cm}^2 \mathrm{g}^{-1} \bigg( \frac{\rho}{10^{-4} \; \mathrm{g} \; \mathrm{cm}^{-3}} \bigg)^{0.5}\\ \bigg( \frac{T}{2500 \; \mathrm{K}} \bigg)^{7.5} \bigg( \frac{Z}{0.02} \bigg)^1.
\end{multline}

The only influence dust has on the H$^-$ opacity is via the metallicity dependence $Z$ of the gas. We set $Z$ equal to the local dust-to-gas ratio in our gas accretion calculations as the metals delivered via dust are present in the gas phase at the H$_2$ dissociation front. Equating $Z$ to the dust-to-gas ratio is justified because the $Z$ dependence of $\kappa (\mathrm{H}^-)$ results from its dependence on the availability of free electrons, most of which are sourced from metallic species. Although some of these metals might be present in the gas, the dust contribution dominates. This is likely to be true even in the most dust depleted regions of the outer disk as CO is predicted to be the dominant gas phase metal in this region. This molecule does not dissociate until much deeper in the planetary atmosphere, and hence it will not contribute free electrons in the region where H$^-$ opacity becomes important.

We use this information to calculate gas accretion rates onto a planetary core as a function of disk location and time using the analytical scaling laws provided by \cite{Lee2015}, modified for the linear dependence on the bound radius and the weak dependence on nebular density \citep[see][]{Lee2020}. The gas-to-core mass ratio (GCR) at time $t$ (with accretion beginning at $t_{\rm 0}$) in the `dusty' planetary envelope regime is given by:

\begin{multline}
    \mathrm{GCR} \sim 0.06 \; f_{\mathrm{R}} \; 
    \bigg( \frac{\Sigma_g}{2000 \; \mathrm{g \; cm^{-3}}} \bigg)^{0.12}
    \bigg( \frac{t - t_{\rm 0}}{1 \; \mathrm{Myr}} \bigg)^{0.4}
    \bigg( \frac{\nabla_{\mathrm{ad}}}{0.17} \bigg)^{3.4} \\ 
    \bigg( \frac{2500 \; \mathrm{K}}{T_{\mathrm{rcb}}} \bigg)^{4.8}
    \bigg( \frac{0.02}{Z} \bigg)^{0.4} 
    \bigg( \frac{\mu_{\mathrm{rcb}}}{2.37} \bigg)^{3.4} \bigg(\frac{M_{\mathrm{core}}}{5 M_{\oplus}} \bigg)^{1.7}.
    \label{eq:gcr}
\end{multline}
Here, $f_{\mathrm{R}}$ is the bounded radius of a planet as a fraction of its min($R_{\mathrm{Hill}}$, $R_{\mathrm{Bondi}}$) and we set it equal to 0.2 \citep[e.g.][]{Fung2019}. The updated scaling law provided by \cite{Lee2020} also allows us to incorporate the dependence of GCR on the gas surface density $\Sigma_g$, which we obtain from our disk model. The normalization factor of 0.06 is valid for $\Sigma_g < 0.1 \times$ MMEN at 0.1 au. $\nabla_{\mathrm{ad}}$, $T_{\mathrm{rcb}}$ and $\mu_{\mathrm{rcb}}$ are the adiabatic gradient, temperature, and the mean molecular weight evaluated at the RCB. We assume a fixed value of $T_{\mathrm{rcb}} = 2500$ K and $\nabla_{\mathrm{ad}} = 0.17$, appropriate for the innermost RCB at the H$_2$ dissociation front, for all our calculations. We calculate $\mu_{\mathrm{rcb}}$ assuming a $\mu = 2.3$ for a pure hydrogen-helium mixture (solar abundance ratio) and $\mu = 17$ for a pure metal-rich atmosphere. For the most metal-rich gases \citep[$Z \gtrsim 0.2$;][]{Lee2016}, the strong dependence of GCR on $\mu_{\mathrm{rcb}}$ dominates over the metallicity-dependent increase in opacity, allowing for rapid accretion \citep{Venturini2015}. Our models predict that the dust-to-gas ratio throughout the disk will remain below this critical value for a majority of the disk lifetime. $Z > 0.2$ in the inner disk only at very early stages ($< 0.1$ Myr) when core formation is still likely ongoing.\footnote{We note that late-stage pollution of an envelope by ambient solids could enhance the interior metallicity beyond $Z \sim 0.2$ and trigger rapid gas accretion \citep{Hori2011}. The short dynamical timescale in the inner disk suggests that the solids there most likely lock into planetary cores before the late-stage disk gas dispersal, and so such late-stage pollution is more likely to occur in the outer disk.} For the entirety of the duration of gas accretion that we model (0.1--10 Myrs), an increased $Z$ therefore acts to reduce the accretion rate by increasing the gas opacity at the RCB. We incorporate the time dependence of $\Sigma_g$, $Z$, and $\mu_{\mathrm{rcb}}$ in our calculation of GCR by numerically differentiating Equation~\ref{eq:gcr} with respect to time and integrating between $t_0 = 0.1$ Myr (the emergence of the core) and time $t$ (in the range $1 - 10$ Myr) at which the planet stops accreting.

\begin{figure}
    \centering
    \includegraphics[width=\linewidth]{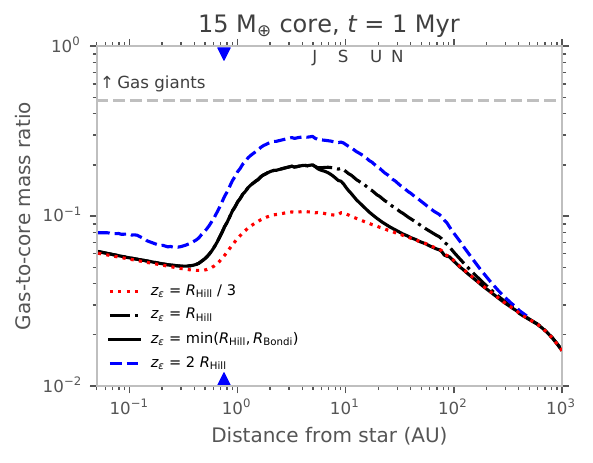}
    \caption{The gas-to-core mass ratio (GCR) at $t = 1$ Myr for a 15 M$_{\oplus}$ core as a function of distance from the star, assuming the core starts accreting at $t_{\rm 0} = 0.1$ Myr. Here, we vary the height $z_{\epsilon}$ from which gas is accreted by the planet. A GCR of 0.48 is marked with a dashed grey line, indicating the threshold for the onset of runaway gas accretion \citep{Lee2014}. The water ice line is marked with blue triangles.}
    \label{fig:gcr_fiducial}
\end{figure}

\begin{figure*}
    \centering
    \includegraphics[width=\linewidth]{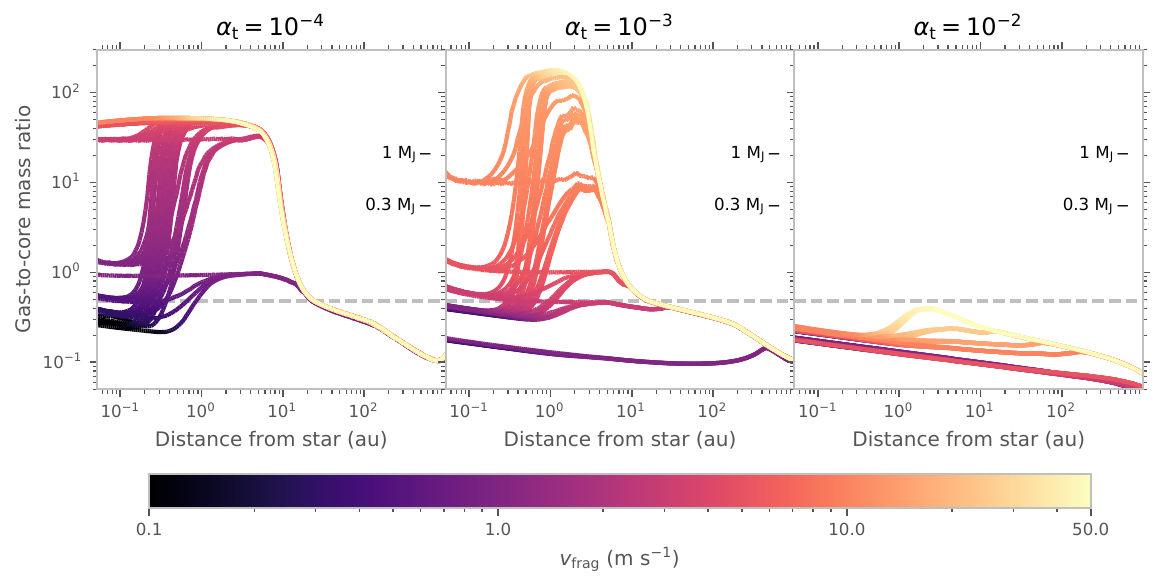}
    \caption{The gas-to-core mass ratio (GCR) at $t = 10$ Myr for a 15 M$_{\oplus}$ core as a function of distance from the star, assuming the core starts accreting at $t_{\rm 0} = 1$ Myr, for our grid of models. We use a later $t_{\rm 0}$ here because the dust-to-gas ratio for the $\alpha_{\rm t} = 10^{-4}$ model varies rapidly at earlier times, and our simple gas accretion model would therefore not be applicable. The dashed grey line indicates the threshold for the onset of runaway gas accretion \citep[GCR of 0.48;][]{Lee2014}. GCRs corresponding to planet masses of $0.3 M_{\rm J}$ and $1 M_{\rm J}$ are marked with black dashes; $0.3 M_{\rm J}$ is commonly used as a lower mass limit when calculating giant planet occurrence rates in RV surveys \citep[e.g.][]{Cumming2008, Wittenmyer2020}.}
    \label{fig:GCR_grid}
\end{figure*}

If a core reaches the threshold for runaway gas accretion (GCR = 0.48, \citealp{Lee2014}), we calculate the subsequent gas accretion rate using the minimum of the cooling-limited rate $\dot{M}_{\rm cool}$, the disk accretion rate $\dot{M}_{\rm disk}$, and the hydrodynamic accretion rate $\dot{M}_{\rm hydro}$ (see \citealp{Lee2019} for an example):
\begin{subequations}
\begin{equation}
    \dot{M}_{\rm cool} = 0.48 \, \frac{M_{\rm core}}{t_{\rm run}} \, {\rm exp} \bigg(\frac{t}{t_{\rm run}}\bigg),
\end{equation}
\begin{equation}
    \dot{M}_{\rm disk} = 3 \, \pi \, \nu \, \Sigma_g,
\end{equation}
\begin{equation}
    \dot{M}_{\rm hydro} = 0.29 \bigg(\frac{M_p}{M_*}\bigg)^{4/3} \frac{\Sigma_g}{1 + 0.034 \, K} \bigg(\frac{r}{H_{\rm gas}}\bigg)^2 r^2 \Omega_K.
\end{equation}
\end{subequations}
Here, $t_{\rm run}$ is the time taken by the core to reach a GCR = 0.48 ($t$ at which GCR = 0.48 minus $t_{\rm 0}$), $M_p = ({\rm GCR} + 1) M_{\rm core}$ is the total planet mass, and $K = (M_p / M_*)^2 \alpha_{\rm t}^{-1} (H_{\rm gas} / r)^{-5}$ accounts for the depletion of gas surface density in the vicinity of the planet due to gap opening. This effect is only included in the gas accretion calculation and not in the evolution of the disk.

Figure~\ref{fig:gcr_fiducial} shows the gas-to-core mass ratio (GCR) calculated for our fiducial core mass of 15 M$_{\oplus}$ as a function of distance from the star at $t = 1$ Myr. We vary the height $z_\epsilon$ from which material is accreted by the planet, which affects the metallicity (dust-to-gas ratio) of the accreted material and therefore the GCR profile. Along with our default value of $z_\epsilon = \mathrm{min}(R_{\mathrm{Hill}}, R_{\mathrm{Bondi}})$, we also show GCR profiles for $z_\epsilon = [1/3, 1, 2] \times R_{\mathrm{Hill}}$. Inside $\sim 1$ au, the relatively high $Z$ ($\sim 0.1$) produces a GCR in the range $0.06 - 0.08$ for a wide range of $z_\epsilon$. However, the sharp drop in $Z$ beyond $\sim 1$ au (see middle panel of Figure~\ref{fig:3d_z_comparison}) leads to a rise in the amount of gas accreted by the planetary core, reaching a peak value of $\sim 0.2$ in the $1 - 10$ au region of the disk for $z_\epsilon = \mathrm{min}(R_{\rm Hill}, R_{\rm Bondi})$. Beyond $\sim 10$ au, the metallicity of the gas (i.e. dust-to-gas ratio) at $R_{\mathrm{Hill}}$ and $R_{\mathrm{Bondi}}$ rises again as the Hill and Bondi radii shrink relative to the disk scale height, which leads to a decline in GCR. The weak dependence of GCR on $\Sigma_g$ also contributes to a decline in GCR with distance. We note that the peak GCR value in the intermediate $1-10$ au region increases with the height above the midplane from which the planet accretes as the dust-to-gas ratio is a strongly decreasing function of $z$ in this region. Overall, Figure~\ref{fig:gcr_fiducial} demonstrates that the amount of gas accreted by a planetary core during the accretion-by-cooling phase, and hence its ability to reach the threshold for runaway growth, varies significantly as a function of its location in the disk.

Figure~\ref{fig:GCR_grid} shows how the final GCR for a 15 M$_{\oplus}$ core varies as a function of the assumed \alphat and \vfrag values. We calculate the GCR assuming that the core begins accreting at $t_{\rm 0} = 1$ Myr, as the dust-to-gas ratio at earlier times varies too rapidly for most of the models with $\alpha_{\rm t} = 10^{-4}$ to allow us to use our simple gas accretion model. We also note that the high dust-to-gas ratio at $t \lesssim 1$ Myr and short orbital timescales in the inner disk for low \vfrag and $\alpha_{\rm t} = 10^{-4}$  (Figure~\ref{fig:grid_plot}) would probably lead to efficient planetesimal formation, thereby reducing the dust-to-gas ratio of the material that is available for gas accretion. GCR calculations for these models at early times therefore require a more careful study of how the dust-to-gas ratio evolves if planetesimal formation occurs, which is beyond the scope of this work.

\begin{figure*}
    \centering
    \includegraphics[width=\linewidth]{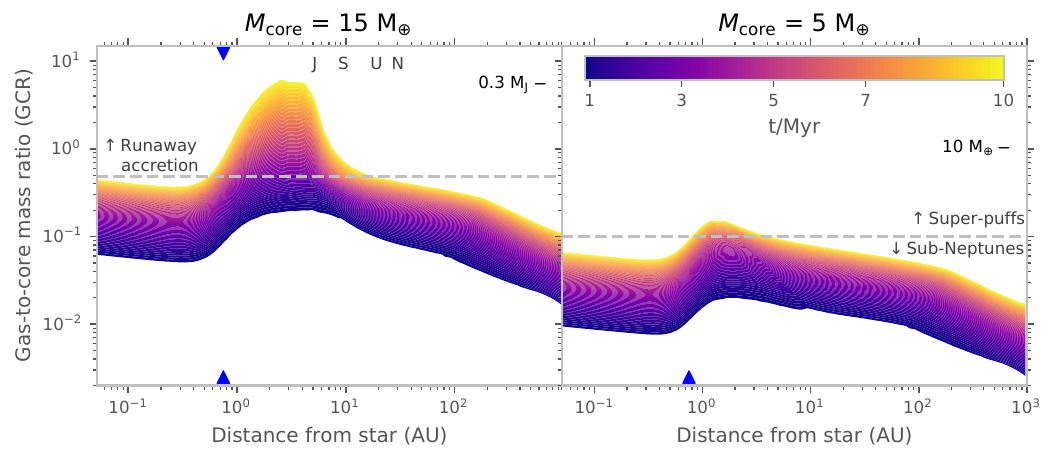}
    \caption{The gas-to-core mass ratio (GCR) as a function of distance from the star for a 15 M$_{\oplus}$ (left panel) and 5 M$_{\oplus}$ (right panel) cores for time $t$ in the range $1-10$ Myr, assuming they start accreting material present at min($R_{\mathrm{Hill}}$, $R_{\mathrm{Bondi}}$) at $t_{\rm 0} = 0.1$ Myr. GCR of 0.48 (onset of runaway gas accretion \citealp{Lee2014}) and GCR = 0.1 (for sub-Neptunes and super-puffs) are marked with dashed grey lines in the top and bottom panels respectively. GCRs corresponding to total planet masses of $0.3 M_{\rm J}$ (left panel) and $10 M_{\oplus}$ (right panel) are indicated as well. The locations of solar system giant planets are marked along the abscissa at the top. The water ice line is marked with blue triangles.}
    \label{fig:GCR_time_core_var}
\end{figure*}

We find that for $\alpha_{\rm t} = 10^{-2}$, the dust-to-gas ratio does not vary significantly with either location or time, and as a result the GCR profiles lie in a relatively narrow range.  The declining GCR with increasing distance is primarily due to the decline in gas surface density. In these high viscosity models, a 15 M$_{\oplus}$ core is unable to reach the threshold for runaway accretion anywhere in the disk. In contrast, models with $\alpha_{\rm t} = 10^{-4}$ and $10^{-3}$ can produce a wide range of GCR values depending on the assumed values of \vfrag. Models with substantial changes in \vfrag across the water ice line result in a large radial variation in the GCR, with a sharp rise at $\sim$1 au and a sharp fall at $\sim$10 au, much more drastic than our fiducial model (Figure~\ref{fig:gcr_fiducial} \& \ref{fig:GCR_time_core_var}). 
These models are characterized by a deep minimum in dust-to-gas ratio at $\sim$1--10 au which accelerates thermodynamic gas accretion, driving planets to runaway whose mass growth is eventually limited by the global disk accretion $\dot{M}_{\rm disk}$ (when GCR $\gtrsim$ 10).


\subsection{Consequences for giant planet formation and demographics}

Our calculations provide a natural explanation for the observed peak in the gas giant planet occurrence rate at $\sim$1--10 au as measured by radial velocity and direct imaging surveys \citep[e.g.][]{Baron2019, Fernandes2019, Nielsen2019, Wittenmyer2020, Fulton2021}. Figure~\ref{fig:GCR_time_core_var} demonstrates that the location of the most favorable sites for rapid gas accretion is driven by the decrease in dust-to-gas ratio just beyond the ice line where relatively larger grains undergo efficient radial drift and vertical settling. We note that the nucleation of gas giants requires relatively massive cores ($\sim$15$M_\oplus$) that assemble early (i.e., accrete gas for at least 3--10 Myrs). Lighter cores and/or those that assemble late (i.e., accrete gas for shorter amount of time) necessarily grow into planets with less massive envelopes. Although it is difficult to obtain good observational constraints on the core masses of extrasolar Jupiters \citep{Thorngren2019}, we note that the cores of sub-Saturns---planets that were on the verge of runaway but were halted in growth before they became gas giants---are better-constrained and appear to range between $\sim$15--20$M_\oplus$ in the limiting case where all metals are assumed to be sequestered in the core \citep{Petigura2017, Lopez2014}. This range also agrees with core mass estimates derived from fitting mass loss models to sub-Saturn occurrence rate as a function of orbital period \citep{Hallatt2021}.

The same change in fragmentation velocity of grains across the ice line that we invoke in our model may also result in the formation of more massive cores outside the ice line \citep[e.g.][]{Morbidelli2015, Venturini2020}, reinforcing our results that gas giants are more likely to originate farther away from the star. Our work further demonstrates that the dust-to-gas ratio is expected to be radially-variant and that it reaches a local minimum at a specific range of orbital distances (1--10 au), creating a preferred zone of rapid gas accretion. Qualitatively, our solar system also fits into our picture, with gas giants Jupiter and Saturn forming at intermediate distances where the GCR peaks and Uranus and Neptune forming further out where the GCR declines with distance \citep{Morbidelli2007, Batygin2010}.

\subsection{Formation of sub-Neptunes and super-puffs}
\label{sec:super_earth_puffs}

Close-in sub-Neptunes appear to possess primordial hydrogen-rich envelopes that are a few percent of the total planet mass \citep[e.g.,][]{Lopez2014, Wolfgang2015, Ning2018}. Given their estimated core masses of $4 - 8$ M$_{\oplus}$ \citep{Wu2019, Rogers2020}, it is difficult to explain why these planets did not undergo runaway gas accretion and turn into gas giants assuming they formed in MMEN and accreted solar metallicity gas. Previous studies have proposed three potential solutions: 1) accretion of metal rich gas, which increases the envelope opacity and slows the gas accretion rate during the cooling growth phase \citep[e.g.][]{Lee2014, Chen2020}, 2) late-time core assembly, so that there is a very short period for the planet to accrete prior to the dispersal of the gas disk \citep{Lee2016}, and 3) a flow of high entropy gas into the Hill sphere of the growing planet that prevents it from cooling (\citealp{Ormel2015, Bethune2019}, but see \citealp{Kurokawa2018}). Scenario 1 in and of itself applies for either dusty or dust-free accretion but it is more effective for dusty accretion as its overall higher opacity delays accretion even more. Our work revisits the first scenario in the context of in situ, dusty gas accretion.

Our results suggest that the enhanced dust-to-gas ratio in the inner disk is sufficient to limit the envelope masses/accretion rates of sub-Neptunes forming in this region. We find that for a representative 5 M$_{\oplus}$ core, the enhanced dust-to-gas ratio inside the ice line is enough to prevent the accretion of a massive gas envelope (Figure~\ref{fig:GCR_time_core_var}, right panel). If the metallicity is too high ($Z > 0.2$), the enhancement in the mean molecular weight of the gas can expedite gas accretion \citep{Lee2015, Venturini2015}. For our fiducial choice of fragmentation velocities and turbulence parameter (as well as for a large swath of the parameter space), $Z$ stays below 0.2 in the inner disk after 0.1 Myrs. As shown in Figure \ref{fig:GCR_time_core_var}, a 5$M_\oplus$ core inside 1 au attains a few percent by mass envelope, consistent with the measured masses and radii of sub-Neptunes, even if the core assembled early and accreted gas for the full 10 Myrs. We note that this result is not mutually exclusive with late-time core assembly for sub-Neptunes. The late-time, gas-poor environment favors the build-up of $\sim$5$M_\oplus$ sub-Neptune cores by a series of collisional mergers. Such mergers are necessary as the isolation masses, either from planetesimal (see \citealp{Dawson2018}, their Figure 2) or pebble accretion (see, e.g., \citealp{Bitsch2018, Fung2018}) are on the order an Earth mass or smaller in the inner disk. Furthermore, late-time assembly of sub-Neptunes prevents inward migration of these planets once they assemble \citep{Lee2016}.

Our models also provide support for previously published hypotheses about the origin of `super-puffs', a rare class of planets with giant planet like radii (4--8 $R_\oplus$) and super-Earth like masses (2--5 $M_\oplus$) \citep{Lee2016}. The low bulk densities of these planets imply that they possess hydrogen-rich envelopes that are tens of \% by mass \citep{Masuda2014, Jontof-Hutter2014, Ofir2014}. Although the gas mass fraction of some super-puffs may be overestimated due to the inflation of planetary radii measurements by photochemical hazes lofted by outflowing gas, the majority of super-puffs do appear to have accreted substantially more gas than sub-Neptunes \citep{Wang2019, Libby-Roberts2020, Gao2020, Chachan2020}. It is difficult to explain how these planets, which have core masses similar to those of sub-Neptunes, could have accreted an order of magnitude more gas in their present-day locations \citep{Ikoma2012, Lee2016}. \cite{Lee2016} proposed that super-puffs might form by accreting `dust-free' gas (dust opacity lower than gas opacity) beyond $\sim 1$ au. Although the dust opacity in our models is never low enough to qualify as dust-free, we find that this decrease in the dust-to-gas ratio beyond the ice line does indeed lead to significantly higher gas accretion rates and GCRs (Figure~\ref{fig:GCR_time_core_var}). All of the currently known super-puffs are in or near orbital resonances with other planets\footnote{Most super-puffs orbit dim stars, which makes it hard to measure their masses with the radial velocity technique. Their masses have typically been determined by transit timing variations, which by definition require them to be in dynamically interacting multi-planet systems.}, which requires relatively smooth convergent migration \citep[e.g.][]{Cresswell2006}. This is consistent with a scenario in which super-puffs formed beyond $\sim 1$ au and then migrated inward via interactions with the protoplanetary gas disk. As Figure \ref{fig:GCR_time_core_var} shows, the creation of super-puffs require their cores to have assembled early so that the total gas accretion time is longer. The requirement for early stage core assembly is also in agreement with the migratory origin of super-puffs as disk-induced migration requires a gas-rich environment.

\section{Discussion and Conclusions}
\label{sec:conclusion}

In this work we use dust evolution models to demonstrate that the dust opacity and dust-to-gas ratio in protoplanetary disks is expected to be radially and vertically variant, with significant implications for planet formation. This is a result of grain growth and transport, which produce a highly non-uniform dust-to-gas ratio in the disk and generate top heavy size distributions with grains that are orders of magnitude larger than the maximum grain size in the commonly-assumed ISM distribution. We explore the sensitivity of our models to assumptions about the disk turbulence and fragmentation velocities and find that we obtain qualitatively similar results over a wide range of plausible values. 

Models with a substantial difference in \vfrag across the ice line and moderate-to-low turbulence values $\alpha_{\mathrm{t}} \lesssim 10^{-3}$ produce the largest radial variations in dust-to-gas ratio and dust opacity. A large change in \vfrag across the ice line leads to a large difference in the Stokes number St of the largest grains within and beyond the ice line. In the inner disk with smaller St (well-coupled to gas), dust grains pile up radially and mix well vertically. In the outer disk with larger St (more decoupled from gas), dust grains drift in rapidly and settle to the midplane. As a result, the inner disk is characterized by high dust-to-gas ratio that is near constant with height, whereas the outer disk is characterized by lower dust-to-gas ratio that decreases even further away from the midplane.

We use our location-dependent dust-to-gas ratio to calculate gas accretion rates onto planetary cores as a function of distance from the star. If we assume that the growing planet predominately accretes material present at min($R_{\mathrm{Hill}}$, $R_{\mathrm{Bondi}}$) above the midplane, we find that the gas-to-core mass ratio (GCR) is a strong function of its location in the disk. Within the ice line, gas accretion onto the core is suppressed by the high dust-to-gas ratio. At intermediate distance beyond the ice line ($1 - 10$ au in our fiducial model), there is a steep decline in the dust-to-gas ratio, causing the GCR to rise and making it easier for cores to reach the threshold for runaway gas accretion. Beyond this point the dust-to-gas ratio increases again as the growing planet accretes from a region closer to the disk midplane (min($R_{\mathrm{Hill}}$, $R_{\mathrm{Bondi}}$) / $H_{\rm gas}$ declines with distance). We conclude that dust-gas dynamics favor gas giant planet formation at intermediate distances, potentially explaining the peak in the giant planet occurrence rate vs.~orbital distance \citep[e.g.][]{Fulton2021}. Our results also provide support for the hypothesis that super-puffs likely formed beyond the ice line, as the lower dust-to-gas ratio in this region can substantially accelerate their gas accretion rates.

We note that the same models presented in this study could be used to constrain the core mass distribution of gas giant exoplanets by quantifying the fraction of planets that undergo runaway gas accretion as a function of location \citep[e.g.][]{Lee2019}. Previous studies on core formation have argued that a change in \vfrag across the ice line could lead to a significant increase in core masses outside the ice line \citep{Morbidelli2015, Venturini2020}. In a future study we will explore whether the radially-varying dust-to-gas ratio alone is sufficient to reproduce the observed mass-period distribution of gas giant exoplanets, or whether it is also necessary to invoke a radially varying core mass function or large scale migration. These same models could also be used to explore why outer gas giants are commonly accompanied by inner super-Earths \citep{Zhu2018, Bryan2019}.

In this study we have limited ourselves to a single fiducial disk model to show how dust opacity varies with radial distance. However, observations of protoplanetary disks indicate that there is a large variation in disk properties such as the disk mass, size, lifetime, and metallicity as well as the mass and luminosity of protostars \citep{Andrews2018, Long2018, Long2019}. In future studies, we will investigate how dust evolution and gas accretion onto planetary cores depend on these properties and whether the diversity of exoplanets is thus linked to the diversity in disk and stellar properties.

Other potential improvements for these calculations include accounting for the conversion of dust to planetesimals/planetary cores on the dust mass budget (likely to be important in the inner disk for $\alpha_{\rm t} = 10^{-4}$ and low \vfrag) and the effect of planet-disk interaction on dust growth and dynamics. In particular, as planetary cores become massive enough to perturb the gas disk, pressure maxima outside the planet's orbit traps some of the dust. This could affect the local size distribution and radial migration of dust as well as the dust-to-gas ratio of the material accreted by the growing planet \citep{Chen2020}. We expect these effects to be perturbative and more localized in nature and the global dust evolution to broadly follow the picture we have painted in this work. Overall, the radial variation of dust-to-gas ratio and dust opacity have a substantial effect on the ability of planetary cores to accrete gas and should be considered in models of planet formation.

\section*{Acknowledgements}
We are grateful to the referees for providing us with thoughtful suggestions that improved the paper. We are indebted to Dave Stevenson, Yanqin Wu, and Chris Ormel for giving us feedback on this manuscript. Y.C. is grateful to Til Birnstiel for providing the excellent public repositories that this work relies on. Support for this work was provided by NASA through Space Telescope Science Institute grant GO-15138. 

\bibliography{manuscript}
\bibliographystyle{apj}

\end{document}